\documentclass[12pt]{article}
\usepackage{amsmath,amssymb,amsopn,graphics,epsfig,color,cite} 

\def\theequation{\arabic{section}.\arabic{equation}}
\newcommand{\Section}[1]{ 
\refstepcounter{section}
\setcounter{equation}{0} 
\setcounter{subsection}{0} 
\setcounter{subsubsection}{0}
\addcontentsline{toc}{section} 
         {\normalsize\textbf{\thesection.\ #1}} 
\bigskip\bigskip\noindent 
\normalsize\textbf{\thesection.\ #1}\nopagebreak\par\smallskip\nopagebreak} 
\def\thesection{{\normalsize\arabic{section}}} 
\renewcommand{\subsection}[1]{ 
\refstepcounter{subsection}
\addcontentsline{toc}{subsection} 
      {\normalsize\normalfont\textit{\thesubsection. #1}} 
\medskip\medskip\noindent 
\normalsize\normalfont
\textit{\thesubsection. #1}\nopagebreak \par\smallskip\nopagebreak} 
\def\thesubsection{{\normalsize
{\arabic{section}.\arabic{subsection}}}} 
\renewcommand{\subsubsection}[1]{ 
\refstepcounter{subsubsection}
\addcontentsline{toc}{subsubsection} 
      {\normalsize\normalfont\textsl{\thesubsubsection. #1}} 
\medskip\medskip\noindent 
\small\normalfont\textsl{\thesubsubsection. #1}
\nopagebreak \par\smallskip\nopagebreak} 
\def\thesubsubsection{{\small
{\arabic{section}.\arabic{subsection}.\arabic{subsubsection}}}} 
\newcounter{appendice}

\newcommand{\mc}[1]{{\mathcal #1}}

\newcommand{\bb}[1]{{\mathbb #1}}

\definecolor{light}{gray}{.9}

\renewcommand{\tilde}{\widetilde}
\renewcommand{\epsilon}{\varepsilon}

\newcommand{\id}{{1 \mskip -5mu {\rm I}}}

\begin{document}

%
%

\begin{titlepage}
\par\vskip 1cm\vskip 2em

\begin{center}
{\Large {\bf
Non equilibrium current fluctuations  
\\
$\phantom{m}$\\
in stochastic lattice gases}
}
\par
\vskip 2.5em \lineskip .5em
{\large
\begin{tabular}[t]{c}
$\mbox{L.\ Bertini}^{1}, 
\,
\mbox{A.\ De Sole}^{2,3},
\,
\mbox{D.\ Gabrielli}^{4},
\,  
\mbox{G.\ Jona-Lasinio}^{5},
\, 
\mbox{C.\ Landim}^{6,7}
$ 
\\
\end{tabular}
\par
}
\par

\medskip
{\small
\begin{tabular}[t]{ll}
{1} & {\it 
Dipartimento di Matematica, Universit\`a di Roma La Sapienza}\\
&  Piazzale Aldo Moro 2, 00185 Roma, Italy\\
{2} & {\it INDAM, Universit\`a di Roma La Sapienza}\\
&  Piazzale Aldo Moro 2, 00185 Roma, Italy\\
{3} & {\it Department of Mathematics, Harvard University}\\
& Cambridge, MA 02138, USA\\
{4} & {\it Dipartimento di Matematica, Universit\`a dell'Aquila}\\
& Coppito, 67100 L'Aquila, Italy\\
{5} & 
{\it Dipartimento di Fisica and INFN, Universit\`a di Roma ``La Sapienza"}\\
& Piazzale A. Moro 2, 00185 Roma, Italy\\
{6} & {\it IMPA}\\
& Estrada Dona Castorina 110, J. Botanico, 22460 Rio de Janeiro, Brasil \\ 
{7} & {\it  CNRS UMR 6085, Universit\'e de Rouen}\\
& 76128 Mont-Saint-Aignan Cedex, France \\
\end{tabular}
}

\end{center}

\vskip 1 em

\centerline{\bf Abstract} 
\smallskip
We study current fluctuations in lattice gases in the macroscopic
limit extending the dynamic approach for density fluctuations
developed in previous articles.  More precisely, we establish a large
deviation principle for a space-time fluctuation $j$ of the empirical
current with a rate functional $\mc I (j)$.  We then estimate the
probability of a fluctuation of the average current over a large time
interval; this probability can be obtained by solving a variational
problem for the functional $\mc I $.  We discuss several possible
scenarios, interpreted as dynamical phase transitions, for this
variational problem. They actually occur in specific models. We
finally discuss the time reversal properties of $\mc I$ and 
derive a fluctuation relationship akin to the Gallavotti-Cohen
theorem for the entropy production.

\vskip 0.8 em

\vfill


\bigskip\bigskip

\end{titlepage}
\vfill\eject

\Section{Introduction}
\label{sec-1}

\subsection{Thermodynamic functionals for non equilibrium systems}
\label{sec-1.0}

In equilibrium statistical mechanics there is a well defined
relationship, established by Boltzmann, between the probability of a
state and its entropy. This fact was exploited by Einstein to study
thermodynamic fluctuations.  So far it does not exist a theory of
irreversible processes of the same generality as equilibrium
statistical mechanics and presumably it cannot exist. While in
equilibrium the Gibbs distribution provides all the information and no
equation of motion has to be solved, the dynamics plays the major role
in non equilibrium.

When we are out of equilibrium, for example in a stationary state of a
system in contact with two reservoirs, even if the system is in a
local equilibrium state so that it is possible to define the local
thermodynamic variables e.g.\ density or magnetization, it is not
completely clear how to define the thermodynamic potentials like the
entropy or the free energy.  One possibility, adopting the
Boltzmann--Einstein point of view, is to use fluctuation theory to
define their non equilibrium analogs. In fact in this way extensive
functionals can be obtained although not necessarily simply additive
due to the presence of long range correlations which seem to be a
rather generic feature of non equilibrium systems.  This possibility
has been pursued in recent years leading to a considerable number of
interesting results. One can recognize two main lines.
The first, as well exemplified by the work of Derrida, Lebowitz and
Speer \cite{DLS1,DLS2,DLS3}, consists in exact 
calculations in specific models of stochastic lattice gases. 
The second is based on a macroscopic dynamical approach
for Markovian microscopic evolutions, 
of which stochastic lattice gases are a main example, 
that leads to some general variational principles. 
We introduced this approach in \cite{BDGJL1,BDGJL2} and developed it in
\cite{BDGJL3,BDGJL4}.
Both approaches have been very effective and of course give the same
results when a comparison is possible.

Let us recall the Boltzmann--Einstein theory of equilibrium thermodynamic
fluctuations, as described for example in  \cite{LL}. The main principle is
that the probability of a fluctuation in a macroscopic region of fixed
volume $V$  is  
\begin{equation}
\label{1}
P \propto \exp\{V\Delta S / k\}
\end{equation}
where $\Delta S$ is the variation of the specific entropy calculated
along a reversible transformation creating the fluctuation and $k$ is
the Boltzmann constant.  Eq.\ (\ref{1}) was derived by Einstein
\cite{E} simply by inverting the Boltzmann relationship between
entropy and probability.  He considered (\ref{1}) as a
phenomenological definition of the probability of a state.  Einstein
theory refers to fluctuations for equilibrium states, that is for
systems isolated or in contact with reservoirs characterized by the
same chemical potentials. 
When in contact with reservoirs $\Delta S$ is the variation of the
total entropy (system + reservoirs) which for fluctuations of constant
volume and temperature is equal to $-{\Delta F}/{T}$, that is minus
the variation of the free energy of the system divided by the
temperature.

We consider a stationary nonequilibrium state (SNS), namely, due to
external fields and/or different chemical potentials at the
boundaries, there is a flow of physical quantities, such as heat,
electric charge, chemical substances, across the system.  To start
with it is not always clear that a closed macroscopic dynamical
description is possible.  If the system can be described by a
hydrodynamic equation, a fact which can be rigorously established in
stochastic lattice gases, a reasonable goal is to find an explicit
connection between the thermodynamic potentials and the dynamical
macroscopic properties like transport coefficients.  As we discussed
in \cite{BDGJL1,BDGJL2,BDGJL3,BDGJL4}, the study of large fluctuations
provides such a connection. It leads in fact to a dynamical theory of
the free energy which is shown to satisfy a Hamilton-Jacobi equation
in infinitely many variables requiring as input the transport
coefficients.  In the case of homogeneous equilibrium states the
solution of the Hamilton-Jacobi equation is easily found, and the
equilibrium free energy is recovered together with the well known
fluctuation-dissipation relationship, widely used in the physical and
physical-chemical literature. On the other hand in SNS the
Hamilton-Jacobi equation is hard to solve.  There are few
one-dimensional models where it reduces to a non linear ordinary
differential equation which, even if it cannot be solved explicitly,
leads to the important conclusion that the non equilibrium free energy
is a non local functional of the thermodynamic variables.  This
implies that correlations over macroscopic scales are present.  The
existence of long range correlations is probably a generic feature of
SNS and more generally of situations where the dynamics is not
invariant under time reversal \cite{BJ}.  As a consequence if we
divide a system into subsystems the free energy is not necessarily
simply additive.

Besides the definition of thermodynamic potentials, in a dynamical
setting a typical question one may ask is the following: what is the
most probable trajectory followed by the system in the spontaneous
emergence of a fluctuation or in its relaxation to an equilibrium or a
stationary state?  To answer this question one first derives a
generalization of the Boltzmann-Einstein formula from which the most
probable trajectory can be calculated by solving a variational
principle.  The free energy is then related to the logarithm of the
probability of such a trajectory and satisfies the Hamilton-Jacobi
equation associated to this variational principle.
For equilibrium states and small fluctuations an answer to this type of
questions was given by Onsager and Machlup in 1953 \cite{OMA}.  The
Onsager-Machlup theory gives the following result under the assumption
of time reversibility of the microscopic dynamics:
the most probable creation and relaxation trajectories of a
fluctuation are one the time reversal of the other.  
As we show in \cite{BDGJL1,BDGJL2}, for SNS 
the Onsager-Machlup relationship has to be modified in the
following way: the spontaneous emergence of a macroscopic
fluctuation takes place most likely following a trajectory which can
be characterized in terms of the time reversed process.

\subsection{Macroscopic dynamics and large fluctuations}

We consider many-body systems in the limit of infinitely many degrees
of freedom. 
Microscopically we assume that the evolution is described
by a Markov process $X_{\tau}$ which represents the state of the
system at time $\tau$.  This hypothesis probably is not so restrictive
because also the dynamics of Hamiltonian systems interacting with
thermostats finally is reduced to the analysis of a Markov process, 
see e.g.\ \cite{epr}.
To be more precise $X_{\tau}$ represents the set of variables
necessary to specify the state of the microscopic constituents
interacting among themselves and with the reservoirs.  The SNS is
described by a stationary, i.e.\ invariant with respect to time
shifts, probability distribution $P_{st}$ over the trajectories of
$X_{\tau}$. We denote by $\mu$ the invariant measure of the process
$X_\tau$. The measure $\mu$ is a probability on the configuration
space and for each fixed time $\tau$ we have $P_{st}(X_\tau=\omega)=
\mu(\omega)$.

We assume that the system admits a macroscopic description in terms of
density fields which are the local thermodynamic variables $\rho_i$.
The usual macroscopic interpretation of Markovianity
is that the time derivatives of the thermodynamic variables
$\dot\rho_i$ at a given instant of time depend only on the $\rho_i$'s
and the affinities (thermodynamic forces) $\frac {\partial F}{\partial
\rho_i}$ at the same instant, recall that $F$ is the free energy.  As
we discussed in \cite{BDGJL3}, for non equilibrium systems, 
the affinities, defined as the
derivative of the non equilibrium free energy, do not determine the
macroscopic evolution of the variables $\rho_i$. There is an
additional non dissipative term which however does not modify the rate
of approach to the stationary state.

For simplicity of notation we assume that there is only one
thermodynamic variable $\rho$ e.g.\ the local density.  
For conservative systems the evolution of the field
$\rho=\rho(t,u)$, where $t$ and $u$ are the macroscopic time and
space coordinates, is then given by the continuity equation 
\begin{equation}
\label{H1}
\partial_t \rho  \;=\;
\nabla \cdot \Big[ \frac 12 D(\rho)\nabla \rho - \chi(\rho) E \Big]
\;=\; - \nabla\cdot J(\rho)
\end{equation}
where $D(\rho)$ is the diffusion matrix, $\chi(\rho)$ the mobility and
$E$ the external field. Finally the interaction with the reservoirs
appears as boundary conditions to be imposed on solutions of
(\ref{H1}). We shall denote by $\bar\rho=\bar\rho(u)$ the unique stationary
solution of (\ref{H1}), i.e.\ $\bar\rho$ is the typical density profile
for the SNS.

This equation derives from the underlying microscopic dynamics through
an appropriate scaling limit in which the microscopic time and space
coordinates $\tau, x$ are rescaled diffusively: $t=\tau /N^2$, $u=x/N$
where $N$ is the linear size of the system so that the number of
degrees of freedom is proportional to $N^d$.  The hydrodynamic
equation (\ref{H1}) represents a law of large numbers with respect to
the probability measure $P_{st}$ conditioned on an initial state
$X_0$. This conditional probability will be denoted by $P_{X_0}$. The
initial conditions for (\ref{H1}) are determined by $X_0$.  Of course
many microscopic configurations give rise to the same value of
$\rho(0,u)$. In general $\rho=\rho(t,u)$ is the limit of the local
density $\pi_N(X_{\tau})$.

The free energy $F(\rho)$, defined as a functional of the
density profile $\rho=\rho(u)$, gives the asymptotic probability
of density fluctuations for the invariant measure $\mu$. More
precisely 
\begin{equation}
\label{BE2}
\mu \big( \pi_N( X ) \approx \rho \big) \sim  
\exp\big\{ - N^d F(\rho) \big\}
\end{equation}
where $d$ is the dimensionality of the system, $\pi_N(X) \approx \rho$ 
means closeness in some metric and $\sim$ denotes
logarithmic equivalence as $N\to\infty$. 
In the above formula we omitted the dependence on the
temperature since it does not play any role in our analysis; we also 
normalized $F$ so that $F(\bar\rho)=0$. 

In the same way, the behavior of space time fluctuations can be
described as follows. The probability that the
evolution of the random variable $\pi_N(X_\tau)$ deviates from the solution
of the hydrodynamic equation and is close to some trajectory
${\hat{\rho}}(t)$ is exponentially small  and of the form
\begin{equation}
\label{LD}
P_{st}\big( \pi_N(X_{N^2t}) \approx  \hat{\rho}(t), \: 
t\in [t_1, t_2]\big) \sim
\exp\big\{-N^d 
\big[ F(\hat \rho({t_1})) + {\mc F}_{[t_1,t_2]}(\hat{\rho}) 
\big] \big\}
\end{equation}
where $\mc F(\hat{\rho})$ is a functional which vanishes if
${\hat{\rho}}(t)$ is a solution of (\ref{H1}) and $F(\hat
\rho({t_1}))$ is the free energy cost to produce the initial density
profile ${\hat{\rho}}({t_1})$.  Therefore $\mc F(\hat{\rho})$
represents the extra cost necessary to follow the trajectory
${\hat{\rho}}(t)$ in the time interval $[t_1,t_2]$.  Equation
(\ref{LD}) is a dynamical generalization of the Boltzmann-Einstein
formula, we shall refer to it as the \emph{dynamical large deviation
principle} with dynamical\emph{rate functional} $\mc F$.
For stochastic lattice gases, as shown in \cite{BDGJL2},
the functional $\mc F$ can be calculated explicitly.

To determine the most probable trajectory followed by the system in
the spontaneous creation of a fluctuation, we consider the following
physical situation.  The system is macroscopically in the stationary
state $\bar\rho$ at $t=-\infty$ but at $t=0$ we find it in the state
$\rho$.  According to (\ref{LD}) the most probable trajectory is the
one that minimizes $\mc F$ among all trajectories $\hat\rho(t)$
connecting $\bar\rho$ to ${\rho}$ in the time interval
$[-\infty,0]$. As shown in \cite{BDGJL1,BDGJL2} this minimization
problem gives the non equilibrium free energy, i.e.
\begin{equation}
\label{l1}
F(\rho)= \inf_{\hat \rho}  \mc F_{[-\infty,0]}(\hat \rho)
\end{equation}
To this variational principle it is naturally associated a
Hamilton-Jacobi equation which plays a crucial role in the analysis
developed in \cite{BDGJL1,BDGJL2,BDGJL3,BDGJL4}.  We emphasize that
the functional $\mc F$, hence the corresponding Hamilton-Jacobi
equation for $F$, is determined by the macroscopic transport
coefficients $D(\rho)$ and $\chi(\rho)$, which are experimentally
accessible, see e.g.\ \cite{oa}.  We can thus regard (\ref{l1}) as a
far reaching generalization of the fluctuation-dissipation theorem
since it allows to express a static quantity like the free energy in
terms of the dynamical macroscopic features of the system.

\subsection{Current fluctuation and related thermodynamic functionals}

Beside the density, a very important observable is the current flux 
\cite{LLL,DDR2,DDR1,PJSB,PJSB2}. This quantity
gives informations that cannot be recovered from the density because
from a density trajectory we can determine the current trajectory only
up to a divergence free vector field.  We emphasize that this is due to the
loss of information in the passage from the microscopic level to the
macroscopic one.

In the previous paper \cite{BDGJL5} we have introduced a
Boltzmann--Einstein type formula for current fluctuations. This
formula shows that  the asymptotic probability, as the number of degrees of
freedom increases, of observing a current fluctuation $j$ on a
space--time domain $[0,T]\times\Lambda$ can be described by a rate
functional $\mc I_{[0,T]}(j)$. 
In the present paper we develop the approach introduced in
\cite{BDGJL5} and illustrate some relevant applications. 

To discuss the current fluctuations, we introduce a vector-valued
observable $\mc J^N(\{X_\sigma$, $0\le \sigma\le \tau\})$ of the
trajectory $X$ which measures the local net flow of particles.  As for
the density, for stochastic lattice gases, we shall be able to derive
a dynamical large deviations principle for the current. Recall that
$P_{X_0}$ stands for the probability $P_{st}$ conditioned on the
initial state $X_0$.
Given a vector field $j:[0,T]\times \Lambda \to \bb R^d$, we have
\begin{equation}
\label{f1a}
P_{X_0} \big( \mc J^N (X) \approx j (t,u) \big) 
\sim \exp\big\{ - N^d \, \mc I_{[0,T]}(j)\big\}
\end{equation}
where the rate functional is 
\begin{equation}
\label{Ica}
\mc I_{[0,T]}(j)\;=\; \frac 12 \int_0^T \!dt \,
\big\langle [ j - J(\rho)  ], \chi(\rho)^{-1}
[ j - J(\rho) ] \big\rangle
\end{equation}
in which we recall that 
\begin{equation*}
J(\rho) = -\frac 12  D(\rho) \nabla \rho  + \chi(\rho) E\; .
\end{equation*}
Moreover, $\rho=\rho(t,u)$ is obtained by solving the continuity
equation $\partial_t\rho +\nabla\cdot j =0$ with the initial condition
$\rho(0)=\rho_0$ associated to $X_0$. The rate functional vanishes if
$j=J(\rho)$, so that $\rho$ solves \eqref{H1}. This is the law of
large numbers for the observable $\mc J^N$. Note that equation
\eqref{Ica} can be interpreted, in analogy to the classical Ohm's law,
as the total energy dissipated in the time interval $[0,T]$ by the
extra current $j - J(\rho)$.

The functional $\mc I$ describes the fluctuation properties of the
current, the density and all observables related to them, as proved in
Section \ref{sec1}.  Among the many problems we can discuss within
this theory, we study the fluctuations of the time average of the
current $\mc J^N$ over a large time interval.  This is the question
addressed in \cite{bd} in one space dimension by postulating an
``additivity principle'' which relates the fluctuation of the time
averaged current in the whole system to the fluctuations in
subsystems.  We show that the probability of observing a given
divergence free time average fluctuation $J$ can be described by a
functional $\Phi(J)$ which we characterize, in any dimension, in terms
of a variational problem for the functional $\mc I_{[0,T]}$
\begin{equation}
\label{limTa}
\Phi (J)
 = \lim_{T\to\infty} \; \inf_{j} 
\frac 1T \; \mc I_{[0,T]} (j)\;,
\end{equation}
where the infimum is carried over all paths $j=j(t,u)$ having time
average $J$. The static additivity principle postulated in \cite{bd}
gives the correct answer only under additional hypotheses which are
not always satisfied. 
Let us denote by $U$ the functional obtained by restricting the
infimum in \eqref{limTa} to divergence free current paths $j$, i.e.\ 
\begin{equation}
\label{Ua}
U(J) =  \inf_{\rho} \frac 12 
\big\langle [ J - J(\rho)  ], \chi(\rho)^{-1}
[ J - J(\rho) ] \big\rangle
\end{equation} 
where the infimum is carried out over all the density profiles
$\rho=\rho(u)$ satisfying the appropriate boundary conditions. From
\eqref{limTa} and \eqref{Ua} it follows that $\Phi \le U$. 
In one space dimension the functional $U$ is the one introduced in
\cite{bd}. 

There are cases in which $\Phi= U$ and in Subsection
\ref{sec5.1} below we give sufficient conditions on the transport
coefficients $D$, $\chi$ for the coincidence of $\Phi$ and $U$.
On the other hand, while $\Phi$ is always convex the functional $U$ 
may be non convex. In such a case $U(J)$ underestimates the
probability of the fluctuation $J$.  In \cite{BDGJL5} we interpreted 
the lack of convexity of $U$, and more generally the strict inequality
$\Phi<U$,  as a dynamical phase transition.  
In the present paper we investigate in more detail the occurence of
this phenomenon. 
Let us denote by
$U^{**}$ the convex envelope of $U$; then $\Phi \le U^{**}$ and in
Subsection~\ref{sec5.2} we give an example where $U^{**} < U$. 

We shall also consider the fluctuation of the time averaged current
with periodic boundary conditions. In Subsection~\ref{sec5.7} we
discuss the behavior of $U$ and $\Phi$ under appropriate conditions on
the transport coefficient and the external field. In particular we
show that for the Kipnis--Marchioro-Presutti (KMP) model \cite{kmp},
which is defined by a harmonic chain with random exchange of energy
between neighboring oscillators, we have $U(J)= (1/2) J^2/ \chi(m)=
(1/2) J^2/m^2 $, where $m$ is the (conserved) total energy. In
addition we show, for $J$ large enough, $\Phi(J) < U(J)$.  This
inequality is obtained by constructing a suitable travelling wave
current path whose cost is less than $U(J)$.  We mention that, by
using the space-time approach introduced in \cite{BDGJL5}, the
possibility of taking advantage of travelling waves has been first
envisaged by Bodineau and Derrida \cite{b1} for the periodic simple
exclusion process with external field.  We refer to the discussion in
Subsection~\ref{sec5.7} for a comparison between KMP and simple
exclusion models.

We study also the behavior of $\mc I$ and $\Phi$ under time reversal
and derive a fluctuation relationship akin to the Gallavotti-Cohen
theorem for the entropy production \cite{gc,k,ls}. In fact, we prove,
in the present context of lattice gases, that the anti-symmetric part
of $\Phi$ is equal to the power produced by the external field and the
reservoirs independently of the details of the model. From this
relationship we derive a macroscopic version the fluctuation theorem
for the entropy production.

\Section{Microscopic model}
\label{sec0}

As the basic microscopic model we consider a stochastic lattice gas
with a weak external field and particle reservoirs at the boundary.
The process can be informally described as follows. We consider
particles evolving on a finite domain. At each site, independently
from the others, particles wait exponential times at the end of which
one of them jumps to a neighboring site. Superimposed to this dynamics,
at the boundary particles are created and annihilated at exponential
times.  More precisely, let $\Lambda\subset\bb R^d$ be a smooth domain
and set $\Lambda_N = N\Lambda\cap \bb Z^d$. We consider a Markov
process on the state space $X^{\Lambda_N}$, where $X$ is a subset of
$\bb N$, e.g.\ $X=\{0,1\}$ when an exclusion principle is imposed.
The number of particles at site $x\in\Lambda_N$ is denoted by $\eta_x
\in X$ and the whole configuration by $\eta\in X^{\Lambda_N}$.  The
dynamics is specified by a continuous time Markov process on the state
space $X^{\Lambda_N}$ with infinitesimal generator $L_N = N^2 \big[
L_{0,N} + L_{b,N} \big]$ defined as follows: for functions
$f:X^{\Lambda_N}\to \bb R$, 
\begin{equation} 
\label{gen}
\begin{array}{rcl} {\displaystyle L_{0,N} f(\eta) } &=& {\displaystyle
\frac 12 \sum_{\substack{x,y\in\Lambda_N\\  |x-y| =1}} c_{x,y}(\eta)
\big[ f(\sigma^{x,y}\eta) - f(\eta) \big] \;, } \\ {\displaystyle
L_{b,N} f(\eta)} &=& {\displaystyle \frac 12
\sum_{\substack{x\in\Lambda_N, y\not\in\Lambda_N\\ |x-y|=1}}
\Big\{ c_{x,y}(\eta) \big[ f(\sigma^{x,y}\eta) - f(\eta) \big] +
c_{y,x}(\eta) \big[ f(\sigma^{y,x}\eta) - f(\eta) \big] \Big\}\;.  }
\end{array} 
\end{equation} 
Here $|x|$ stands for the usual Euclidean norm and $\sigma^{x,y}\eta$,
$x,y\in\Lambda_N$, for the configuration obtained from $\eta$ by
moving a particle from $x$ to $y$:
\begin{equation*}
\big( \sigma^{x,y}\eta \big)_z = \left\{
\begin{array}{lcl} \eta_z & \textrm{ if } & z\neq x,y \\ \eta_y +1 &
\textrm{ if } & z= y \\ \eta_x -1 & \textrm{ if } & z= x\;.
\end{array} \right.  
\end{equation*} 
If $x\in\Lambda_N$, $y\not\in\Lambda_N$, then $\sigma^{y,x}\eta$ is
obtained from $\eta$ by creating a particle at $x$, while
$\sigma^{x,y}\eta$ is obtained by annihilating a particle at $x$.
Therefore the generator $L_{0,N}$ describes the bulk dynamics which
preserves the total number of particles whereas $L_{b,N}$ models the
particle reservoirs at the boundary of $\Lambda_N$. Note that we
already speeded up the microscopic time by $N^2$ in the definition of
$L_N$, which corresponds to the diffusive scaling.

Assume that the bulk rates $c_{x,y}$, $x,y\in \Lambda_N$, satisfy the
local detailed balance \cite{ls} with respect to a Gibbs measure
defined by a Hamiltonian $\mc H$ and in presence of an external vector
field $E=(E_1,\dots,E_d)$ smooth on the macroscopic scale.  Likewise,
assume that the boundary rates $c_{x,y}$, $c_{y,x}$, $x\in\Lambda_N$,
$y\not\in\Lambda_N$, satisfy the local detailed balance with respect
to $\mc H$ and in presence of a chemical potential $\lambda_0(y/N)$
smooth on the macroscopic scale.

The above requirements are met by the following formal definitions.
Fix a smooth function $\lambda_0 : \Lambda\to \bb R$ and a Hamiltonian
$\mc H$. Consider jump rates $c^0_{x,y}$ satisfying the detailed
balance with respect to the Gibbs measure associated to $\mc H$ with
free boundary conditions if $x,y\in\Lambda_N$, while if $x\in
\Lambda_N$, $y\not\in \Lambda_N$ we add the chemical potential
$\lambda_0(y/N)$:
\begin{equation*}
\begin{array}{l}
\vphantom{\Big\{}
{\displaystyle 
c_{x,y}^0(\eta) \;=\;
\exp - \big\{ \mc H(\sigma^{x,y}\eta) - \mc H(\eta) \big\}
\, c_{y,x}^0 (\sigma^{x,y} \eta)\,,
\quad x,y\in\Lambda_N \,;      
}
\\
\vphantom{\Big\{}
{\displaystyle 
c_{x,y}^0(\eta) \;=\;
\exp - \big\{ \mc H(\sigma^{x,y}\eta) - \mc H(\eta) + \lambda_0(y/N) \big\}
\, c_{y,x}^0 (\sigma^{x,y} \eta)\,,
\quad x\in\Lambda_N,\, y\not\in \Lambda_N\;.
}
\end{array}
\end{equation*}
Note that we included the inverse temperature in the Hamiltonian $\mc
H$. Of course if $\eta_x=0$ then $c^0_{x,y}(\eta) =0$.

Fix a smooth vector field $E = (E_1, \dots, E_d):\Lambda \to\bb R^d$
and let
\begin{equation}
\label{ratas}
c_{x,x+e_i}(\eta) \;:=\; e^{ N^{-1}  E_i(x/N)} 
\, c_{x,x+e_i}^0(\eta) \;,\quad
c_{x+e_i,x}(\eta) \;:=\; e^{ - N^{-1}  E_i(x/N)} 
\, c_{x+e_i,x}^0(\eta) \;, 
\end{equation}
where $\{e_1, \dots, e_d\}$ stands for the canonical basis in $\bb
R^d$. Namely, for $N$ large, by expanding the exponential, particles
at site $x$ feel a drift $N^{-1} E(x/N)$.

Typically, for a non equilibrium model, we would consider $\Lambda$ as
the $d$-dimensional cube of side one, the system under a constant
field $E/N$ and a chemical potential $\lambda_0$ satisfying
$\lambda_0(y/N) = \gamma_0$ if the first coordinate of $y$ is $0$,
$\lambda_0(y/N) = \gamma_1$ if the first coordinate of $y$ is $N$,
imposing periodic boundary conditions in the other directions of
$\Lambda$.

By setting $c_{x,y}=0$ if both $x$ and $y$ do not belong to
$\Lambda_N$, we can rewrite the full generator $L_N$ as follows
\begin{equation}
\label{genmc}
L_{N} f(\eta) 
= \frac {N^2}2 \sum_{\substack{x,y \in \bb Z^d \\ |x-y|=1}}
c_{x,y}(\eta) \big[ f(\sigma^{x,y}\eta) - f(\eta) \big] 
\end{equation}

We consider an initial condition $\eta\in X^{\Lambda_N}$. The
trajectory of the Markov process $\eta(t)$, $t\ge 0$, is an element on
the path space $D\big(\bb R_+;X^{\Lambda_N}\big)$, which consists of
piecewise constant paths with values in $X^{\Lambda_N}$. We shall
denote by ${\bb P}_{\eta}^N$ the probability measure on $D\big(\bb
R_+;X^{\Lambda_N}\big)$ corresponding to the distribution of the
process $\eta(t)$, $t\ge 0$ with initial condition $\eta$. 

Examples of stochastic lattices gases are the simple exclusion
processes in which $X=\{0,1\}$, $\mc H =0$ and $c^0_{x,y} (\eta) =
\eta_x[1-\eta_y]$ and zero range processes in which $X= \bb N$, $\mc H
(\eta) = \sum_x \sum_{1\le k\le \eta(x)} \log g(k)$, for some function
$g: \bb N \to\bb R_+$ such that $g(0)=0$, and $c^0_{x,y} (\eta) =
g(\eta_x)$.

\Section{Macroscopic description of lattice gases}
\label{sec1}

The empirical density $\pi^N$ can be naturally defined as follows.
To each microscopic configuration $\eta\in X^{\Lambda_N}$ we associate
a macroscopic profile $\pi^N(u)$, $u\in \Lambda$, by requiring that 
for any smooth function $G:\Lambda \to  \bb R$ 
\begin{equation}
\label{emden}
\langle \pi^N, G \rangle = 
\int_{\Lambda} \!du\: \pi^N(u) \, G(u) 
= \frac{1}{N^{d}} \sum_{x\in\Lambda_N} G(x/N) \eta_x 
\end{equation}
so that $\pi^N(u)$ is the local density at the macroscopic point
$u=x/N$ in $\Lambda$. Of course $\pi^N(u)$ is really a sum of point
masses at the points $x/N$ with weight $\eta_x/N^d$; in the limit
$N\to\infty$ it will however weakly converge to a ``true'' function
$\rho(u)$.

The definition of the empirical current is slightly more
complicated. Indeed it is not a function of the configuration $\eta\in
X^{\Lambda_N}$ but of the trajectory $\{\eta(t)\}_{t\ge 0} \in D(\bb
R_+;X^{\Lambda_N})$.  Given an oriented bond $(x,y)$, let $\mc
N^{x,y}(t)$ be the number of particles that jumped from $x$ to $y$ in
the time interval $[0,t]$.  Here we adopt the convention that $\mc
N^{x,y}(t)$ is the number of particles created at $y$ due to the
reservoir at $x$ if $x\not\in \Lambda_N$, $y\in\Lambda_N$ and that
$\mc N^{x,y}(t)$ represents the number of particles that left the
system at $x$ by jumping to $y$ if $x\in \Lambda_N$, $y\not
\in\Lambda_N$.  The difference $Q^{x,y}(t)= \mc N^{x,y}(t) - \mc
N^{y,x}(t)$ is the net number of particles flown across the bond
$\{x,y\}$ in the time interval $[0,t]$.  Given a trajectory $\eta(s)$,
$0\le s\le t$, the instantaneous current across $\{x,y\}$ is defined
as $d Q^{x,y}_{t}/ dt $.  This is a sum of $\delta$--functions
localized at the jump times with weight $+1$, resp.\ $-1$, if a
particle jumped from $x$ to $y$, resp.\ from $y$ to $x$.

For a given realization of the process $\eta(t)$ in 
$D\big(\bb R_+;X^{\Lambda_N}\big)$,  we define the corresponding  
empirical current $\mc J^N$ as follows.
Let $T>0$ and pick a smooth vector field  $G=(G_1, \dots,G_d)$ defined on
$[0,T]\times\Lambda$. We then set
\begin{eqnarray}
\label{empcurr}
\langle\!\langle 
\mc J^N, G \rangle\!\rangle_T 
&=& \int_0^T\!dt \int _{\Lambda} du \, G(t,u) \cdot \mc J^N(t,u)  
\nonumber \\
&=& \frac{1}{N^{d+1}} \sum_{i=1}^d  \sum_{x} \int_0^T  G_i(t, x/N) \,
d Q^{x,x+e_i} (t)\;,
\end{eqnarray}
where $\cdot$ stands for the inner product in $\bb R^d$ and we sum
over all $x$ such that either $x\in \Lambda_N$ or $x+e_i\in
\Lambda_N$.  The empirical current $\mc J^N$ is therefore a signed
measure on $\big( [0,T]\times\Lambda \big)^d$, while we recall that
the empirical density is a positive measure on $\Lambda$.  The
normalization $N^{-(d+1)}$ in (\ref{empcurr}) has been chosen so that
the empirical current has a finite limit as $N\to\infty$.

The local conservation of the number of particles is expressed by 
$$
\eta_x(t) - \eta_x(0) +\sum_{y: |x-y|=1} Q^{x,y}(t) =0\;.
$$
It gives the following continuity equation for the 
empirical density and current. 
Let $G$ be a smooth function on a neighborhood of the closure of 
$\Lambda$. Denote by $\nabla_N G$ the vector field whose 
coordinates are $\big(\nabla_N G\big)_i(u) = 
N [G( u+e_i/N)-G(u)]$. Then  
$$
\langle \pi^N(T), G \rangle - \langle \pi^N(0), G \rangle
= 
\langle\!\langle \mc J^N , \nabla_N G \rangle\!\rangle_T 
- 
\int_0^T 
\frac{1}{N^d} 
\sum_{ \substack{x\in\Lambda_N, y \not\in \Lambda_N \\ |x-y|=1} }
G(y/N) dQ^{x,y}(t)
$$
The above equation can be formally stated as the 
continuity equation
\begin{equation}
\label{contN}
\partial_t \pi^N + \nabla_N \cdot \mc J^N = 0
\end{equation}
In particular, given the initial condition, 
the trajectory of the process, described by the empirical density 
$\pi^N$ can be completely recovered from the empirical current $\mc
J^N$.  
\medskip

We briefly discuss at the heuristic level the law of large numbers, as
$N\to\infty$, for the empirical density and the empirical
current. Details are given in Appendix \ref{secap}.  Fix a sequence of
configurations $\eta^N$ and assume that its associated empirical
measure $\pi^N$ converges to $\rho_0 (u) du$ for some density profile
$\rho_0 : \Lambda \to\bb R_+$.  Let us denote by $\rho=\rho(t,u)$,
$J=J(t,u)$, the limiting values of $\pi^N(t,u)$, $\mc J^N(t,u)$,
respectively.  Here $\pi^N(t,u)$ is the empirical density associated
to the configuration $\eta(t)$ and $\mc J^N(t,u)$ has been defined in
(\ref{empcurr}).
 
The microscopic relation (\ref{contN}) implies the continuity equation 
\begin{equation}
\label{cont}
\partial_t \rho + \nabla\cdot J = 0
\end{equation}
To derive a closed evolution for $\rho$ and $J$, we need to express
the current $J$ in terms of the density $\rho$. To simplify the
exposition, we assume the process to be gradient: there
exist local functions $h_0^{(i)}(\eta)$, $i=1,\dots,d$,  
depending on the configuration $\eta$ around $0$, so that for any
$i=1,\cdots, d$
\begin{equation*}
c^0_{x,x+e_i}(\eta) - c^0_{x+e_i,x}(\eta)
= h^{(i)}_{x} (\eta) - h^{(i)}_{x+e_i} (\eta) 
\end{equation*}
where $h^{(i)}_x$ is the function $h_0^{(i)}$ evaluated on the
configuration $\eta$ translated by $x$.

Denote by $\mu_{\lambda}$ the infinite volume grand canonical ensemble
relative to the Hamiltonian $\mc H$ with chemical potential
$\lambda$. Choose the chemical potential $\lambda = \lambda(\rho)$ so
that $\mu_\lambda[\eta_0] = \rho$ and define
\begin{equation}
\label{dc}
d^{(i)}(\rho) \:=\;
\mu_{\lambda(\rho)}  \big[ h^{(i)}_0 \big] \;,\quad 
\chi^{(i)}(\rho) \;:=\; (1/2) 
\mu_{\lambda(\rho)}  \big[ c^0_{0,e_i} + c^0_{e_i,0} \big] 
\end{equation}
We show in Appendix \ref{secap} that the current $J$ can be expressed
in terms of the density $\rho$ as
\begin{equation}
\label{f8}
J = -\frac 12  D(\rho) \nabla \rho  + \chi(\rho) E =: J(\rho)
\end{equation}
where $D$ and $\chi$ are $d\times d$ diagonal matrices with entries
$D_{ii}(\rho) = \frac{d}{d\rho} d^{(i)} (\rho)$ and $\chi_{ii}(\rho) =
\chi^{(i)} (\rho)$. 

For non gradient systems the diffusion matrix $D$ and the mobility
$\chi$ are not in general diagonal. In such a situation $D$ is given
by a Green--Kubo formula \cite[II.2.2]{sp} and $\chi$ can be obtained
by linear response theory \cite[II.2.5]{sp}. These coefficients are
related by Einstein relation $D=R^{-1}\chi$, where $R$ is the
compressibility: $R^{-1} = F_0''$, in which $F_0$ is the equilibrium
free energy associated to the Hamiltonian $\mc H$, \cite{sp}.

To conclude the description of the evolution, it remains to examine
the evolution at the boundary of $\Lambda$. We claim that the density
is fixed there because we speeded up diffusively the non-conservative
Glauber dynamics at the boundary:
\begin{equation}
\label{3.14}
\lambda\big(\rho(t,u)\big) = \lambda_0(u)
\quad\quad u\in\partial\Lambda
\end{equation}

The macroscopic evolution of the density and the current is thus
described by the equation
\begin{equation*}
\left\{
\begin{array}{l}
\partial_t \rho + \nabla\cdot J = 0 \;, \quad u\in\Lambda \\
J = -\frac 12  D(\rho) \nabla \rho  + \chi(\rho) E\;, 
\quad u\in\Lambda \\
\lambda\big(\rho(t,u)\big) = \lambda_0(u) \;,
\quad u\in\partial\Lambda \\
\rho(0,\cdot) = \rho_0(\cdot)\; .
\end{array}
\right.
\end{equation*}

The stationary density profile $\bar\rho=\bar\rho(u)$, $u\in\Lambda$,
is the stationary solution of the hydrodynamic equation, that is
\begin{equation*}
\left\{
\begin{array}{l}
\vphantom{\Big(}
{\displaystyle 
\nabla\cdot J(\bar\rho(u)) = 0 \;, \quad u\in\Lambda
}
\\ 
\vphantom{\Big(}
{\displaystyle 
\lambda\big(\bar\rho(u)\big) = \lambda_0(u)
\quad u \in\partial\Lambda\,,
}
\end{array}
\right.
\end{equation*}
If we let the macroscopic time diverge, $t\to \infty$, $\rho(t) \to
\bar\rho$ and $J(\rho(t))$ converges to $J(\bar\rho)$, which is the
current maintained by the stationary state.
\medskip 

We next discuss the large deviation properties of the empirical
current. More details are given in Appendix \ref{secap}. As before we
consider a sequence of initial configuration $\eta^N$ such that the
empirical density $\pi^N(\eta^N)$ converges to some density profile
$\rho_0$. We fix a smooth vector field $j:[0,T]\times \Lambda \to \bb
R^d$.  The large deviation principle for the current states that
\begin{equation}
\label{f1}
\bb P_{\eta^N}^N \big( \mc J^N (t,u) \approx j (t,u) ,\; 
(t,u)\in [0,T]\times\Lambda \big) 
\sim \exp\big\{ - N^d \, \mc I_{[0,T]}(j)\big\}
\end{equation}
where the rate functional $\mc I$ is 
\begin{equation}
\label{Ic}
\mc I_{[0,T]}(j)\;=\; \frac 12 \int_0^T \!dt \,
\big\langle [ j(t) - J(\rho(t))  ], \chi(\rho(t))^{-1}
[ j(t) - J(\rho(t)) ] \big\rangle
\end{equation}
in which $\rho(t)=\rho(t,u)$ is obtained by solving the continuity
equation
\begin{equation}
\label{ce}
\left\{
\begin{array}{l}
\partial_t\rho(t,u) +\nabla\cdot j(t,u) =0
\\
\rho(0,u)= \rho_0(u)
\end{array}
\right.
\end{equation}
and $J(\rho)$ is given by (\ref{f8}).

Of course there are compatibility conditions to be satisfied, for
instance if we have chosen a $j$ such that $\rho(t,u)$ becomes
negative for some $(t,u)\in [0,T]\times\Lambda$ then $\mc I_{[0,T]}(j)
= +\infty$. Notice that, even if not indicated explicitly in the
notation, the rate functional $\mc I$ depends on the initial density
profile $\rho_0$, through equation (\ref{ce}).

We note that in the large deviation functional (\ref{Ic}) the
fluctuation of the density $\rho(t)$ is determined by the current
$j(t)$. The large deviations properties of the density, which we
described in \cite{BDGJL1,BDGJL2,BDGJL4} for non equilibrium
stochastic lattice gases, can thus be deduced from the ones of the
current, see Appendix~\ref{secap} for the details.  This is due to the
fact that the continuity equation, as already remarked, holds exactly
at the microscopic level, see (\ref{contN}).  On the other hand the
constitutive equation (\ref{f8}) holds only in the limit $N\to\infty$
when fluctuations can be neglected.

\Section{Large deviation of the time averaged current}
\label{s:tac}

We want to study the fluctuations of the time average of the empirical
current over a large time interval $[0,T]$; the corresponding
probability can be obtained from the space time large deviation
principle (\ref{f1}). 
Fix $T>0$ and a divergence free vector field $J=J(u)$. We introduce
the set of possible paths $j$ of the current with time average $J$ 
$$
\mc A_{T,J} = \Big\{ j = j(t,u) \, :\:
\frac 1T \int_0^T \!dt \: j(t,u) = J(u)
\Big\}
$$ 
The condition of vanishing divergence on $J$ is required by the
local conservation of the number of particles.  By the large
deviations principle (\ref{f1}), for $T$ and $N$ large we have
\begin{eqnarray}
\label{LT}
\bb P_{\eta^N}^N \Big( \frac 1T \int_0^T \!dt \; \mc J^N (t) \approx J 
\Big) \sim  \exp \big\{-N^d T \Phi (J) \big\}
\end{eqnarray}
where the logarithmic equivalence is understood by sending {\em
first\/} $N\to\infty$ and {\em then\/} $T\to \infty$.  In Subsection
\ref{sec4} below we shall show that for the zero range process the
limits can be taken in the opposite order; we expect this to be true
in general.  The functional $\Phi$ is given by
\begin{equation}
\label{limT}
\Phi (J)
 \; = \; \lim_{T\to\infty} \; \inf_{j\in \mc A_{T,J}} 
\frac 1T \; \mc I_{[0,T]} (j)
\; = \;\inf_{T>0} \; \inf_{j\in \mc A_{T,J}} 
\frac 1T \; \mc I_{[0,T]} (j)
\end{equation}
By a standard sub--additivity argument we show that 
the limit $T\to\infty$ exists and coincides with the infimum in $T$. 
Indeed, given $j_1 \in \mc A_{T,J}$ and  $j_2 \in \mc A_{S,J}$, we
have 
\begin{equation}
\label{numerino}
\mc I_{[0,T+S]} (j)  = \mc I_{[0,T]} (j_1) +  \mc I_{[0,S]} (j_2) 
\end{equation}
where $j$ is obtained by gluing $j_1$ and $j_2$. Here we used the
invariance of $\mc I$ under time shift and that $j_1 \in \mc A_{T,J}$
implies that the corresponding density $\rho_1$, obtained by solving
the continuity equation (\ref{ce}), satisfies
$\rho_1(0)=\rho_1(T)$. From the previous equation we get the
sub--additivity property:
$$
\inf_{j\in \mc A_{T+S,J}}  \mc I_{[0,T+S]} (j)
\le \inf_{j\in \mc A_{T,J}}  \mc I_{[0,T]} (j)
+ \inf_{j\in \mc A_{S,J}}  \mc I_{[0,S]} (j)
$$ 
Even if the rate functional $\mc I$ depends on the initial density
profile $\rho_0$, by taking the limit in (\ref{limT}) it is easy to
show $\Phi$ does not.

We now prove that $\Phi$ is a convex functional.  Let $0<p<1$ and 
$J=pJ_1+(1-p)J_2$, we want to show that $\Phi(J)\leq
p\Phi(J_1)+(1-p)\Phi(J_2)$.  By (\ref{limT}), given $\epsilon>0$ we
can find $T>0$, $j_1 \in \mc A_{pT,J_1}$, and $j_2 \in \mc
A_{(1-p)T,J_2}$ so that
$$
\begin{array}{lcl}
{\displaystyle 
\Phi(J_1) }
&\ge &
{\displaystyle 
\frac 1{pT} \: \mc I_{[0,pT]} (j_1) -\epsilon
}\\
{\displaystyle 
\Phi(J_2) }
&\ge &
{\displaystyle 
\frac 1{(1-p)T} \: \mc I_{[0,(1-p)T]} (j_2) -\epsilon
}
\end{array}
$$ 
By the same arguments used in (\ref{numerino}), the path obtained
by gluing $j_1$ with $j_2$, denoted by $j$, is in the set $\mc
A_{T,J}$. Therefore,
$$
\Phi(J) \; \le \; \frac 1T \: \mc I_{[0,T]} (j) 
\; \le \;  p \, \Phi(J_1) + (1-p) \, \Phi(J_2) + \epsilon  
$$ 
which proves the convexity of $\Phi$.  These arguments are standard
in proving the existence and convexity of thermodynamic functions in
statistical mechanics.

\medskip
We next study the variational problem on the right hand side of 
(\ref{limT}). We begin by deriving an upper bound. 
Given $\rho=\rho(u)$ and  $J=J(u)$, $\nabla\cdot J=0$,
let us introduce the functionals
\begin{eqnarray}
\label{cU}
\mc U (\rho,J) &=& \frac12 \langle J - J(\rho), \chi(\rho)^{-1} 
[J - J(\rho)] \rangle
\\
\label{U}
U(J)& = &\inf_\rho \; \mc U(\rho, J)
\end{eqnarray}
where the minimum in (\ref{U}) is carried over all profiles $\rho$
satisfying the boundary condition (\ref{3.14}) 
and $J(\rho)$ is given by (\ref{f8}).
When $J$ is constant, that is, in the one--dimensional case, 
the functional $U$ is the one introduced in \cite{bd}.

We claim that
\begin{equation}
\label{1ub}
\Phi(J) \le U(J)\;.
\end{equation}
The strategy to prove this bound is quite simple, see also \cite{bd}.
Let $\hat\rho=\hat\rho(J)$ be the density profile which minimizes the
variational problem (\ref{U}).  Given the initial density profile
$\rho_0$, we choose some fixed time $\tau>0$ and a current $\hat\jmath
= \hat\jmath(u)$ which moves the density from $\rho_0$ to $\hat\rho$
in a time lag $\tau$, namely such that $\tau \nabla \cdot\hat\jmath =
\rho_0 -\hat\rho$.  We now construct the path $j=j(t,u)$, $(t,u)\in
[0,T]\times\Lambda$ as follows
\begin{equation*}
j(t) = \left\{
\begin{array}{ccl}
\hat \jmath &\textrm{if} & 0\le t < \tau \\
\vphantom{\Big\{}
\frac{T}{T -2 \tau} \: J &\textrm{if} & \tau \le t < T -\tau \\
- \hat \jmath &\textrm{if} & T -\tau  \le t \le T 
\end{array}
\right.
\end{equation*}
The corresponding density $\rho(t)$ is obtained by solving the
continuity equation (\ref{ce}), i.e.\ 
\begin{equation*}
\rho (t) = \left\{
\begin{array}{ccl}
\rho_0 + \frac{t}{\tau} (\hat\rho -\rho_0) 
&\textrm{if} & 0\le t < \tau \\
\vphantom{\Big\{}
\hat\rho   &\textrm{if} & \tau \le t < T -\tau \\
\rho_0 + \frac{T-t}{\tau} (\hat\rho -\rho_0) 
&\textrm{if} & T -\tau  \le t \le T 
\end{array}
\right.
\end{equation*}
It is straightforward to verify that $j\in \mc A_{T, J}$, 
as well as $\lim_{T\to\infty} \frac{1}{T} \: \mc I_{[0,T]}(j) = U(J)$.

By the convexity of $\Phi(J)$ we can improve the 
upper bound (\ref{1ub}) for free.
Let us denote by  $U^{**}$ the convex envelope of $U$, i.e.\ the largest
convex functional below $U$. By taking the convex envelope in (\ref{1ub}) 
we get
\begin{equation}
\label{ub}
\Phi(J) \le U^{**}(J)
\end{equation}

\smallskip
We next discuss a lower bound for the variational problem (\ref{limT}). 
We denote by ${\tilde {\mc U}}$ and  ${\tilde U}$ the same functionals
as in (\ref{cU})--(\ref{U}), but now defined on the space of all
currents without the conditions of vanishing divergence. 
Let also ${\tilde U}^{**}$ be the convex envelope of $\tilde U$. 
Let  $j\in\mc A_{T,J}$. By the convexity of ${\tilde {U}}^{**}$
in the set of all currents, we get
$$
\begin{array}{lcl}
{\displaystyle
\frac 1T \mc I_{[0,T]}(j)  
}
&= & 
{\displaystyle
\frac 1T \int_0^T\!dt  \: {\tilde {\mc U}} (\rho(t),j(t))
\; \ge \; 
\frac 1T \int_0^T\!dt  \: {\tilde U} (j(t))
}
\\
&\ge &
{\displaystyle
\frac 1T \int_0^T\!dt  \: {\tilde U}^{**} (j(t))
\; \ge \; {\tilde U}^{**}(J)
}
\end{array}
$$
which implies
\begin{equation}
\label{lb}
\Phi(J) \; \ge \; {\tilde U}^{**}(J)
\end{equation}
The upper and lower bounds (\ref{ub}) and (\ref{lb}) are different in
general. For a divergence free $J$ we have ${\tilde U}(J)= {U}(J)$ but
since the convex envelopes are considered in different spaces, we only
have ${\tilde U}^{**}(J) \le {U}^{**}(J)$.

The derivation of the upper bound shows that our result differs from
the one in \cite{bd} if $U$ is not convex. Moreover, if $\Phi (J) <
U(J)$, the optimal density path $\rho$ in the variational problem
\eqref{limT} must be time dependent.

\medskip
We now examine how different behaviors of the solution to the
variational problem (\ref{limT}) reflect different dynamical regimes
that we interpret as dynamical phase transitions.  It is convenient to
work in the time interval $[-T,T]$ instead of $[0,T]$.  We consider
the system in the ensemble defined by conditioning on the event $
(2T)^{-1} \int_{-T}^T\!dt \: \mc J^N(t)= J$ with $N$ and $T$ large.
The parameter $J$ plays therefore the role of an intensive
thermodynamic variable and the convexity of $\Phi$ expresses a
stability property with respect to variations of $J$.

If $\Phi(J)=U(J)$ and the minimum for (\ref{U}) is attained for
$\rho=\hat\rho(J)$ we have a state analogous to a unique phase: by
observing the system at any fixed time $t$ we see, with probability
converging to one as $N,T\to \infty$, the density $\pi^N(t) \sim
\hat\rho(J)$ and the current $ \mc J^N(t)\sim J$.

When $\Phi(J) =U^{**}(J) <U(J) $, we have a state analogous to a phase 
coexistence. Suppose for example $J= p J_1 + (1-p)J_2$ 
and $U(J) > U^{**} (J)= p U(J_1) + (1-p) U(J_2)$ for some
$p,J_1,J_2$.  The values $p,J_1,J_2$ are determined by $J$ and $U$.
The density profile is then not determined, but rather we observe with
probability $p$ the profile $\hat\rho(J_1)$ and with probability $1-p$
the profile $\hat\rho(J_2)$. Actually there is a memory
of initial condition: if we take $\rho(-T)=\rho(T)=\hat\rho(J_1)$
we will see the density $\hat\rho(J_1)$ in the time intervals 
$[-T,-(1-p)T]$ and $[(1-p)T,T]$, $\hat\rho(J_2)$ in $[-(1-p)T,(1-p)T]$.

Consider now the case in which a minimizer for (\ref{limT}) is a
function $\hat \jmath (t)$ not constant in $t$. This is possible (an
example will be given in Subsection~\ref{sec5.7}) only when 
$\Phi(J) < U^{**}(J)$.  Suppose
first that $\hat \jmath (t)$ is periodic with period $\tau$ and denote
by $\hat\rho(t)$ the corresponding density. Of course we have
$\tau^{-1}\int_0^\tau\!dt \: \hat\jmath(t) = J$. In such a case we
have in fact a one parameter family of minimizers which are obtained
by a time shift $\alpha\in[0,\tau]$.  By choosing $2T$ an integral
multiple of $\tau$ and $\rho(-T)=\hat\rho(\alpha)$ for some
$\alpha\in[0,\tau]$ then the empirical density in the conditional
ensemble will follow the path $\hat\rho(t+\alpha+T)$. This behavior
is analogous to a non translation invariant state in equilibrium
statistical mechanics, like a crystal. Finally if $\hat \jmath (t)$ is
time dependent and not periodic the corresponding state is analogous
to a quasi--crystal.

\medskip
The asymptotic \eqref{LT} can be formulated in terms of the Laplace
transform of the empirical current as follows. For each divergence
free, time independent, vector field $\lambda=\lambda(u)$ we have
\begin{equation}
\label{fa1}
\lim_{T\to\infty}\lim_{N\to \infty}
\frac{1}{T \,N^d} \log 
\bb E^N_{\eta^N} \Big( 
e^{ N^d \langle\!\langle \mc J^N,\lambda \rangle\!\rangle_T}
\Big) = \Phi^* (\lambda)
\end{equation}
where $\Phi^* (\lambda)$ is the Legendre transform of $\Phi (J)$:
\begin{equation*}
\Phi^* (\lambda) = \sup_{J} \big\{ \langle \lambda, J\rangle - \Phi (J)
\big\}\;,
\end{equation*}
where the supremum is carried over all the divergence free vector
fields $J$.  It follows from \eqref{1ub} that $U^* \le \Phi^*$.
\medskip

We conclude this section deriving a variational expression for $U^*$.
Recall the definitions \eqref{cU}, \eqref{U} of $U$.
\begin{equation*}
\begin{array}{rcl}
{\displaystyle
\! U^* (\lambda) 
}
&\!=\!&
{\displaystyle 
\sup_{J,\rho} \Big\{ 
\langle \lambda, J\rangle 
-\frac 12 \langle [ J-J(\rho)] ,\chi(\rho)^{-1}[ J-J(\rho)] \rangle
\Big\}
}
\\
&\!=\!&
{\displaystyle 
\sup_{J,\rho} \Big\{
-\frac 12 \langle [ J-J(\rho)-\chi(\rho) \lambda] ,
\chi(\rho)^{-1} [ J-J(\rho)-\chi(\rho)\lambda] \rangle
}
\\
&&
{\displaystyle
 \phantom{\sup_{J,\rho} \Big\{ }
\qquad\qquad\qquad\qquad\qquad\qquad\qquad\qquad\qquad  
+\frac 12 \langle \lambda, \chi(\rho)\lambda \rangle 
+\langle \lambda, J(\rho)\rangle 
\Big\}
}
\end{array}
\end{equation*}
To compute the supremum over $J$ we decompose the vector field 
$J(\rho)+\chi(\rho)\lambda$ as follows
\begin{equation}
\label{decomp}
J(\rho)+\chi(\rho)\lambda = \chi(\rho)\nabla \psi 
+ \big[ J(\rho)+\chi(\rho) \big(\lambda  -\nabla \psi\big)\big]
\end{equation}
where $\psi$ solves
\begin{equation*}
\left\{
\begin{array}{ll}
{\displaystyle
\nabla\cdot\big( \chi(\rho) \nabla \psi \big) 
= \nabla\cdot( J(\rho) +\chi(\rho)\lambda \big)
}
& \; u\in\Lambda
\\
{\displaystyle \vphantom{\Big\{}
\psi(u) = 0  
}
& \; u\in\partial\Lambda
\end{array}
\right.
\end{equation*}
Since the second term in the decomposition (\ref{decomp}) is
divergence free we get
\begin{equation}
\label{Uh}
U^*(\lambda) \;=\; \sup_{\rho} \Big\{
-\frac 12 \langle \nabla \psi, \chi(\rho) \nabla \psi \rangle
+\frac 12 \langle \lambda, \chi(\rho) \lambda \rangle 
+\langle \lambda, J(\rho)\rangle 
\Big\}
\end{equation}
where the supremum is over all density profiles $\rho$ satisfying
$F_0'(\rho(u)) = \lambda_0(u)$, $u\in\partial \Lambda$.

\Section{Time--reversal and Gallavotti--Cohen symmetry}

In this Section we discuss the properties of the rate functional for
the current under time reversal. We also show that the functional
$\Phi$, which measures the probability of deviations of the time
averaged current, satisfies a fluctuation theorem analogous to the 
Gallavotti--Cohen symmetry.

\subsection{Time--reversal properties of the rate functional}

In the previous Sections we discussed a large deviation principle
given a fixed initial condition $\eta^N$ associated to a density
profile $\rho_0$, i.e.\ $\pi^N(\eta^N) \to \rho_0$.  Now we consider
instead the stationary process, namely the initial condition is
distributed according to the invariant measure $\mu^N$ which is
defined by $\sum_{\eta} \mu^N(\eta) L_N f(\eta) = 0$ for any
observable $f:X^{\Lambda_N}\to \bb R$; recall the generator $L_N$ has
been defined in \eqref{genmc}. As discussed in the Introduction, the
large deviations of the empirical density under the distribution
$\mu^N$ are described by the non equilibrium free energy $F$, i.e.,
\begin{equation}
\label{S}
\mu^N \big( \pi^N \approx \rho \big) \sim \exp\big\{ - N^d F(\rho)
\big\}
\end{equation}
In \cite{BDGJL1,BDGJL2} we show that the functional $F$, which for
equilibrium states is trivially related to the free energy, can be
characterized by a variational problem on the dynamical rate
functional for the density $\mc F$ introduced in (\ref{LD}), see also 
(\ref{rfden}). To this variational problem is associated a
Hamilton--Jacobi equation which plays a crucial role.

In order to analyze the large deviations properties of the stationary
process, since the initial condition is not fixed, it is natural to
consider the joint fluctuations of the empirical density and
current. We have
\begin{equation}
\label{LDjoint}
\bb P_{\mu^N}^N \big( \pi^N \approx \rho, \mc J^N\approx j \;\; t\in [-T,T]
\big) 
\sim \exp\{- N^d {\mathcal G}_{[-T,T]} (\rho,j)\}
\end{equation}
Here $\bb P_{\mu^N}^N$, a probability measure on the space
$D\big( \bb R; X^{\Lambda_N} \big)$, is the stationary process.
Of course, fluctuations of the density and of the current are not
independent since the continuity equation $\partial_t{\rho}+\nabla\cdot j
=0$ must be satisfied. Therefore the large deviation functional is
\begin{equation}
\label{fjoint} 
{\mathcal G}_{[-T,T]} (\rho,j) 
= \left\{
\begin{array}{ll}
{\displaystyle 
F(\rho(-T))+{\mathcal I}_{[-T,T]}(j) 
}
& \textrm{if }  
\partial_t {\rho}+\nabla \cdot j=0 
\\
{\displaystyle \vphantom{\Big\{}}
+\infty & \textrm{otherwise}  \\
\end{array}
\right.
\end{equation}
where $\mc I$ has been introduced in (\ref{Ic}).
If we are interested only in the current fluctuations in the
stationary process we get the appropriate rate functional by
projecting (\ref{fjoint}), 
\begin{equation*}
\inf_{\rho_0}\left\{F(\rho_0)+{\mathcal I}_{[-T,T]}(j)\right\}\;.
\end{equation*}

Let us denote by $L_N^a$ the adjoint of the generator $L_N$
(\ref{genmc}) with respect to the invariant measure $\mu^N$. We call the
process generated by $L_N^a$, which is still Markovian, 
the adjoint process. We remark that the invariant measure of the
adjoint process is again $\mu^N$.
Given a path $\eta \in D\big( \bb R; X^{\Lambda_N} \big)$ its time
reversed is naturally defined as $[\vartheta \eta] (t) = \eta (-t)$.
The stationary adjoint process, that we denote by $\bb
P_{\mu^N}^{N,a}$, is the time reversal of $\bb P_{\mu^N}^{N}$, i.e.\ 
we have $\bb P_{\mu^N}^{N,a} =\bb P_{\mu^N}^{N} \circ \vartheta^{-1}$.
We extend the definition of the time reversal operator $\vartheta$ 
to the current as $[\vartheta j](t)=-j(-t)$. Note that the current $j$
changes sign under time--reversal. 
Then
\begin{equation*}
\bb P^{N}_{\mu^N} \Big( \pi^N \approx \rho,\: \mc J^N \approx j
\;\; t\in [-T,T] \Big) 
 =  \bb P^{N,a}_{\mu^N} \Big( \pi^N \approx \vartheta\rho, 
\: \mc J^N \approx \vartheta j  \;\; t \in [-T,T]\Big)
\end{equation*}
At the level of large deviations this implies
\begin{equation}
\label{GGa}
{\mathcal G}_{[-T,T]}(\rho,j)={\mathcal 
G}^a_{[-T,T]}(\vartheta\rho,\vartheta j) 
\end{equation}
where ${\mathcal G}^a_{[-T,T]}$ is the large deviation functional
for the adjoint process. 

The relationship (\ref{GGa}) has far reaching consequences. We next
show that it implies a fluctuation dissipation relation for the
current. We assume that the adjoint process has a dynamical large
deviations principle 
of the same form as (\ref{fjoint}) with ${\mathcal I}$ replaced by 
${\mathcal I}^a$ where
\begin{equation*}
\mc I^a_{[-T,T]}(j)\;=\; \frac 12 \int_{-T}^T \!dt \, \langle [ j(t)
- J^a(\rho(t))  ], \chi(\rho(t))^{-1} [ j(t) - J^a(\rho(t)) ] \rangle\;,
\end{equation*}
in which $J^a(\rho)$ is the typical value of the current of the
adjoint process. 
We divide both sides of (\ref{GGa}) by $2T$ and take the limit $T\to
0$.  By using (\ref{fjoint}) we get
\begin{equation*}
\Big\langle \frac{\delta F}{\delta\rho}, \partial_t \rho 
\Big\rangle
= \frac 12
\big\langle j - J(\rho), \chi(\rho)^{-1} [j - J(\rho) ] \big\rangle
- \frac 12
\big\langle j + J^a(\rho), \chi(\rho)^{-1} [j + J^a(\rho) ] 
\big\rangle
\end{equation*}
recalling that $\partial_t \rho +\nabla \cdot j =0$, this is
equivalent to 
\begin{equation*}
-\Big\langle \frac{\delta F}{\delta\rho}, \nabla \cdot j 
\Big\rangle
= 
-\big\langle J(\rho)+ J^a(\rho), \chi(\rho)^{-1} j \big\rangle
+\frac 12
\big\langle J (\rho)+  J^a(\rho), \chi(\rho)^{-1} [J(\rho)-J^a(\rho) ]
\big\rangle
\end{equation*}
which has to be satisfied for any $\rho$ and $j$.  By using that
$\delta F / \delta \rho$ vanishes at the boundary of $\Lambda$, see
\cite{BDGJL2}, we can integrate by parts the left hand side above and
get
\begin{eqnarray}
\label{j+ja}
&& 
J(\rho)+J^a(\rho)=-\chi(\rho)\nabla\frac{\delta F}{\delta \rho}
\\
&&
\label{hjnuovo}
\big\langle J (\rho), \chi(\rho)^{-1} J(\rho) \big\rangle
=\big\langle J^a (\rho), \chi(\rho)^{-1} J^a(\rho) \big\rangle
\end{eqnarray}

Equation (\ref{j+ja}) is a fluctuation dissipation for the current
analogous to the one for the density discussed in \cite{BDGJL2}. It
also extends the relationships between currents and thermodynamic
forces, see e.g.\ \cite{ONS1}, to a non equilibrium setting.  By
plugging (\ref{j+ja}) into (\ref{hjnuovo}) we also get another
derivation of the Hamilton--Jacobi equation mentioned before, i.e.\
\begin{equation*}
\frac 12 \Big\langle \nabla \frac{\delta F}{\delta \rho}, \chi(\rho) 
\nabla \frac{\delta F}{\delta \rho} \Big\rangle 
+ \Big\langle \frac{\delta F}{\delta \rho}, \nabla \cdot J(\rho)
\Big\rangle = 0
\end{equation*}

\medskip
Let us now consider the variational problem (\ref{limT}) as well as
the same problem for the functional $\mc I^a$, we denote by $\Phi^a$
the corresponding functional. From (\ref{GGa}) we
get
\begin{equation}
\Phi(J)=\Phi^a(-J)
\end{equation}
For reversible process this
symmetry states that the functional $\Phi $ is even.

Let us consider a path $j(t)$, $t\in [-T,T]$ such that $(2T)^{-1}
\int_{-T}^{T}\!dt \: j(t) = J$ for some divergence free vector field
$J$. Recalling (\ref{f8}) and that
$D(\rho)\chi(\rho)^{-1}=F_0''(\rho)$ we have
$$
\chi(\rho)^{-1} J(\rho) = - \frac12 \nabla F_0'(\rho) + E
$$
Since $F_0'(\rho(u))=\lambda_0(u)$, $u\in \partial \Lambda$, by developing the
square in (\ref{Ic}) and integrating by parts we get 
\begin{equation}
\label{GC1}
\frac{1}{2T} 
\mc G_{[-T,T]}(\rho,j) = 
\frac{1}{2T} \mc G_{[-T,T]}(\vartheta\rho,\vartheta j) 
- 2 \langle J ,E \rangle 
+ \int_{\partial\Lambda} \!d\Sigma \: \lambda_0 \, J \cdot \hat{n} 
\end{equation}
where $d\Sigma$ is the surface measure on $\partial\Lambda$ 
and $\hat{n}$ is the outward normal to $\Lambda$.
In particular this relation implies that if $\hat\rho,\hat\jmath$ is
an optimal path for the variational problem defining $\Phi(J)$ then 
$\vartheta\hat\rho,\vartheta\hat\jmath$ is
an optimal path for the variational problem defining $\Phi(-J)$. 

By taking the limit $T\to \infty$ in (\ref{GC1}) we get
\begin{equation}
\label{GC2}
\Phi(J)- \Phi(-J)=\Phi(J)- \Phi^a(J)=
- 2 \langle J ,E \rangle 
+\int_{\partial\Lambda} \!d\Sigma \: \lambda_0 \, J \cdot \hat{n} 
\end{equation}
which is a Gallavotti--Cohen type symmetry in our space time dependent
setup for macroscopic observables.  Note that the right hand side of
(\ref{GC2}) is the power produced by the external field and the
boundary reservoirs (recall $E$ is the external field and $\lambda_0$
the chemical potential of the boundary reservoirs).

\subsection{Entropy production}

Recall that we denote by $\bb P^N_{\mu^N}$ the stationary state and by 
$\bb P^{N,a}_{\mu^N}$ its time reversed, i.e.\ the stationary adjoint
process. In the context of Markov processes the Gallavotti--Cohen
observable is defined as 
\begin{equation}
\label{fgc}
W_N(T) = - \frac{1}{2T N^d} 
\log \frac {d\bb P_\mu^{N,a}}{d\bb P_\mu^N} \bigg|_{[-T,T]} 
\end{equation}
where the subscript means that we consider both distributions in
the time interval $[-T,T]$. We introduced the factor $2 T \, N^d$ 
in order to discuss the asymptotic $N, T\to\infty$.
As discussed in \cite[\S 2.4]{ls}, $W_N(T)$ can be interpreted as the
microscopic production of the Gibbs entropy. 
For $N$ fixed and $T\to \infty$ the functional $W_N$ satisfies a
large deviation principle with rate function $f_N$ namely,
\begin{equation}
\label{f5}
\bb P^N_{\mu^N} \big(  W_N(T) \approx q \big) 
\; \sim \; \exp\big\{- 2 T  N^d \, f_N(q) \big\}
\end{equation} 
In \cite{gc,k,ls} it is shown that $W_N$ satisfies the Gallavotti--Cohen
symmetry, which states that the odd part of $f_N$ is linear with a
universal coefficient: $f_N(q) - f_N(-q) = - q$.

An elementary computation, analogous to the one in \cite{ls}, shows
that, for the stochastic lattice gases as introduced in Section
\ref{sec0}, we can express the functional $W_N$ in terms of the
empirical current. More precisely, we have
\begin{equation*}
\begin{array}{l}
{\displaystyle 
W_N(T) 
} =  {\displaystyle - 
\frac1{2T N^d}
\bigg\{ 
\log\frac{\mu^N(\eta(T))}{\mu^N(\eta(-T))} 
+ \mc H (\eta(T)) -\mc H (\eta(-T))
}
\\
\quad\quad {\displaystyle \phantom{-\Bigg\{}
- \frac{2}{N} \sum_{j=1}^d \sum_{x}
   E_j (x/N)   Q^{x,x+e_j}([-T,T])  
+ \sum_{\substack{x\in\Lambda_N\\y\not\in\Lambda_N}} 
\lambda_0 (y/N) Q^{x,y}([-T,T])  
\bigg\}\;,
}
\end{array}
\end{equation*}
where the summation is carried over all $x$ such that either $x\in
\Lambda_N$ or $x+e_j\in \Lambda_N$. The previous equation can be
understood as an entropy balance. Indeed, in the right hand side the
first term is, for $N$ large, the difference of the non equilibrium
free energy at times $T$ and $-T$, the second is the difference of the
energy and the third is the work done by the external field $E$
and the boundary reservoirs. Therefore, $W_N$ can be interpreted as the
total entropy produced by the system in the time interval $[-T, T]$.

Recalling the definition of the empirical current $\mc J^N$, we can
rewrite the above equation as
\begin{equation}
\label{wnbis}
\begin{array}{lcl}
{\displaystyle 
W_N(T) 
}
&=&  {\displaystyle \frac{1}{2T} \bigg\{ 
- \frac{1}{N^d} \Big[ \log\frac{\mu^N(\eta(T))}{\mu^N(\eta(-T))}  
+ \mc H (\eta(T)) -\mc H (\eta(-T))
\Big]
}
\\
&& {\displaystyle \phantom{-\Bigg\{}
+ 2 \langle\!\langle \mc J^N, E \rangle\!\rangle_{[-T,T]}
- \frac{1}{N^d} \sum_{\substack{x\in\Lambda_N\\y\not\in\Lambda_N}} 
\lambda_0 (y/N) Q^{x,y}([-T,T])  
\bigg\}
}
\end{array}
\end{equation}
We emphasize that, while the empirical current is a vector in $\bb
R^d$, $W_N(t)$ is a scalar.

From the previous expression it follows that, for any $\delta > 0$
\begin{equation}
\label{llnwn}
\lim_{T\to\infty}\lim_{N\to\infty} 
\bb P_{\eta^N}^N  \Big(
\Big|
  W_N(T) - 
2\langle E, J(\bar\rho)  \rangle  
- \int_{\partial\Lambda}\!d \Sigma \; \lambda_0 
J(\bar\rho) \cdot \hat n  
\Big| > \delta \Big)= 0
\end{equation}
where we recall that 
$J(\bar\rho) = - (1/2) D(\bar\rho)\nabla\bar\rho + \chi(\bar\rho) E$ 
is the typical current.

We note that as $T\to \infty$ we can neglected the first line on the
r.h.s.\ of (\ref{wnbis}) because it is a boundary term. We thus  define
\begin{equation}
\label{wntil}
\widetilde
W_N(T) 
=  \frac{1}{2T} \Big\{ 
2 \langle\!\langle \mc J^N, E \rangle\!\rangle_{[-T,T]}
- \frac{1}{N^d} \sum_{\substack{x\in\Lambda_N\\y\not\in\Lambda_N}} 
\lambda_0 (y/N) Q^{x,y}[-T,T]
\Big\}
\end{equation}
which satisfies, as $T\to\infty$ with $N$ fixed, 
the large deviation estimate (\ref{f5}) with the same rate function $f_N$.

On the other hand, since $\tilde W_N(T)$ is a function of the
empirical current, we can apply the large deviation principle
(\ref{LDjoint}). We then get, by taking \emph{first} the limit  
$N\to\infty$ and \emph{then} $T\to\infty$,
\begin{equation}
\label{f5bis}
\bb P^N_{\mu^N} \big(  \widetilde W_N(T) \approx q \big) 
\; \sim \; \exp\big\{- 2 T  N^d \, f (q) \big\}
\end{equation} 
where the rate function $f$ can be expressed in terms of the 
functional $\mc I$ namely
\begin{equation*}
f (q)= \lim_{T\to\infty}  \inf_{j\in \mc B_{T,q}} 
\frac{1}{2T} \mc I_{[-T,T]} (j)
\end{equation*}
in which we introduced the set of currents
\begin{equation}
\label{atq}
\mc B_{T,q} := \bigg\{ j \, : \:  
\frac{1}{2T} \bigg[ 
2 \int_{-T}^{T}\!dt \: \langle j(t), E \rangle -
\int_{-T}^{T}\!dt \int_{\partial\Lambda}\!d \Sigma \; \lambda_0
j(t)\cdot \hat n  \bigg] = q
\bigg\}
\end{equation}
where we recall $d \Sigma$ is the surface measure on $\partial\Lambda$
and $\hat n$ is the outward normal to $\Lambda$.

Finally, since $E$ and $\lambda_0$ are time independent we can take
the time average of the empirical current in  (\ref{atq}).
Recalling (\ref{limT}), we get 
\begin{equation}
\label{fq}
f (q)= \inf_{J\in \mc B_q} \Phi (J)
\end{equation}
where 
\begin{equation}
\label{aq}
\mc B_q := \Big\{ J \, : \:  
2 \langle J, E \rangle  - 
\int_{\partial\Lambda}\!d \Sigma \; \lambda_0 J \cdot \hat n 
= q 
\,, \; \nabla \cdot J =0 \Big\}
\end{equation}
where we inserted the condition $\nabla \cdot J =0$ because other
things do not happen. 
The content of the variational problem (\ref{fq}) is to look for,
among all possible currents, the best one to have a fixed entropy
production.  
It is straightforward to verify that the symmetry (\ref{GC2}) implies
the classical Gallavotti--Cohen symmetry 
for the limiting functional $f$, i.e.\ $f(q)-f(-q)=-q$. On the other
hand, if $d>1$, equation (\ref{GC2}) is more general than the 
classical Gallavotti--Cohen symmetry.

In the above argument we first took the limit $N\to\infty$ and next
$T\to\infty$, but we expect that these limits could be taken in any
order. In particular these would imply 
$\lim_{N\to\infty} f_N(q) = f(q)$. In Section \ref{sec4} we prove 
that this is the case for the zero range process.
We finally note that in the one--dimensional case, setting
$\Lambda=[0,1]$, we can easily solve (\ref{fq}). We get
$$
f(q)= \Phi \Big( \frac{q}{ 2 \langle E \rangle - 
[\lambda_0(1) -\lambda_0(0)] } \Big)
$$

\Section{Dynamical phase transitions: examples}

As we have discussed  in Section~\ref{s:tac}, we always have the following
inequalities
\begin{equation}
\label{inequalities}
{\tilde U}^{**}(J) \leq \Phi(J) \leq U^{**}(J) \leq U(J)
\end{equation}
for any divergence free $J$.  A natural question is when the above
inequalities are strict and when are equalities, in particular when
$\Phi=U$.  
As discussed in Section~\ref{s:tac}, the strict inequality 
$\Phi(J) < U(J)$ is a dynamical phase transition on the ensemble defined by
conditioning on the event in which the time average current equals $J$.
In this Section we discuss several examples which show that
different scenarios actually do take place in concrete models.
As we have shown in Section~\ref{sec1} the macroscopic behavior
(including the probability of large fluctuations) of the system is
determined by the transport coefficients $D(\rho)$ and
$\chi(\rho)$. In this Section we consider these as given functions
and discuss the properties of the variational problem defining
$\Phi$. Specific choices of $D$ and $\chi$ correspond
to well studied microscopic models, such as the simple exclsion processes,
the zero range process, and the KMP model \cite{kmp}.  

In Subsection~\ref{sec5.1} we find sufficient conditions on 
$D$ and $\chi$ implying $\Phi =U$. 
In Subsection~\ref{sec5.7} we discuss periodic boundary conditions: 
under appropriate conditions on the transport coefficient and the
external field we show that the minimizer for the variational problem
defining $U$ is obtained when $\rho$ is constant (in space). 
Moreover, by considering travelling waves, we find for $J$ large a 
better (space-time dependent) strategy so that $\Phi < U$. 
These conditions hold in particular for the KMP
model with no external field. 
Moreover, for the exclusion process with sufficiently large 
external field, we show that there exists a  travelling wave 
path of current  whose cost is strictly less than the constant (in
time and space) one.  This was first observed in \cite{b1}.
In Subsection~\ref{sec5.2}, we give an example where $U$ is non
convex which implies $\Phi<U$. 
Finally, in Subsection~\ref{sec4} we compute the Legendre transform of
$U$ for the one dimensional zero range process in the presence of
external field. As a by product, we show that the macroscopic limit
$N\uparrow\infty$ and $T\uparrow\infty$ can be interchanged.

\subsection{A sufficient condition for $\Phi=U$}
\label{sec5.1}

We consider the case when the matrices $D(\rho)$ and
$\chi(\rho)$ are multiple of the identity, i.e., there are strictly
positive scalar functions still denoted by $D(\rho)$, $\chi(\rho)$, so
that $D(\rho)_{i,j}= D(\rho) \delta_{i,j}$, $\chi(\rho)_{i,j}=
\chi(\rho) \delta_{i,j}$, $i,j=1,\dots ,d$. We denote derivatives with
a superscript.
Let us first consider the case with no external field, i.e.\ $E=0$, we
shall prove that if 
\begin{equation}
\label{c<}
D(\rho) \chi''(\rho) \le D'(\rho) \chi'(\rho) 
\quad \textrm{ for any } \rho 
\end{equation}
then $\Phi=U$. In this case $U$ is necessarily convex. 

Moreover we show that if  
\begin{equation}
\label{c=}
D(\rho) \chi''(\rho) = D'(\rho) \chi'(\rho) 
\quad \textrm{ for any } \rho 
\end{equation}
then we have $\Phi=U$ for any external field $E$.  We mention that
under the condition (\ref{c=}), as shown in \cite[\S 7]{bgl}, also the
non equilibrium free energy $F$ can be computed explicitly and it
is a local functional.

Condition (\ref{c<}) is satisfied for the symmetric simple exclusion
process, where $D=1$ and $\chi(\rho)=\rho(1-\rho)$,
$\rho\in[0,1]$. Condition (\ref{c=}) is satisfied for the zero range
model, where $D(\rho) = \Psi '(\rho)$ and $\chi(\rho) =\Psi(\rho)$ for
some strictly increasing function $\Psi : \bb R_+ \to \bb R_+$.
Condition (\ref{c=}) is also satisfied for the non interacting
Ginzburg--Landau model, where, $\rho\in\bb R$, $D(\rho)$ is an
arbitrary strictly positive function and $\chi(\rho)$ is constant.

Let us consider first the case when $E=0$ and condition (\ref{c<})
holds. In view of (\ref{Ic}) and (\ref{cU}), to prove that $\Phi=U$ it
is enough to show that for each $j=j(t,u) \in \mc A_{T,J}$, i.e., such
that $T^{-1}\int_0^T\!dt \: j(t) = J$, and $\rho(t)$ such that
$\partial_t \rho(t)+\nabla \cdot j(t)=0$ we have
\begin{equation}
\label{lbc}
\frac{1}{T} {\mc I}_{[0,T]} (j) =
\frac{1}{T} \int_0^T\!dt \: \mc U (\rho(t), j(t)) \ge U(J)
\end{equation}

Instead of $\rho$ we introduce a new variable $\alpha$ so that
$\alpha= d (\rho) := \int^{\rho} \!d\rho' D(\rho')$.  Condition
(\ref{c<}) is then equivalent to the concavity of the function
$X(\alpha):= \chi \big( d^{-1}(\alpha) \big)$ where $d^{-1}$
is the inverse function of $d$. We introduce the functional 
$$
\mc V (\alpha,j) := \mc U ( d^{-1}(\alpha), j)
= \frac 12 
\Big\langle j+ \frac 12 \nabla \alpha,  
 \frac 1{X(\alpha)} 
\big[  j+ \frac 12 \nabla \alpha \big] \Big\rangle
$$
where we used (\ref{f8}).
We claim that the functional $\mc V$ is jointly convex in
$(\alpha,j)$. Let us first show that this implies the lower bound 
\eqref{lbc}. We have 
$$
\begin{array}{l}
{\displaystyle
\frac{1}{T} \int_0^T\!dt \: \mc U (\rho(t), j(t)) 
}
\;=\; 
{\displaystyle
\frac{1}{T} \int_0^T\!dt \: \mc V (\alpha(t), j(t)) 
}
\\
\qquad
\ge \;
{\displaystyle
\vphantom{\Bigg\{}
\mc V \Big( \frac{1}{T} \int_0^T\!dt \:\alpha(t), 
\frac{1}{T} \int_0^T\!dt \: j(t) \Big)
\; \ge \; 
\inf_{\alpha} \mc V (\alpha, J) \; = \;  
\inf_{\rho} \mc U (\rho, J) = U(J)
}
\end{array}
$$
in which we used the convexity of $\mc V$ in the second step and $j\in\mc
A_{T,J}$ in the third.

To prove that $\mc V$ is jointly convex we write
$$
\mc V(\alpha,j)= \sup_{a} \mc V_a( \alpha,j) \,,\qquad
\mc V_a(\alpha,j)
:= \big\langle j+\frac 12 \nabla\alpha , a \big\rangle
-\frac 12 \big\langle a , 
{X(\alpha)} a \big\rangle
$$
and the supremum is taken over all smooth vector fields
$a=a(u)$ on $\Lambda$.
Since $X(\alpha)$ is
concave, for each fixed $a$ the functional $\mc V_a$ is jointly convex.
The claim follows.

In the case with non vanishing $E$, we can use
the same argument, but the functional $\mc V$ is given by 
$$
\mc V (\alpha,j) := \mc U ( d^{-1}(\alpha), j)
= \frac 12 
\Big\langle j+ \frac 12 \nabla \alpha 
-  X(\alpha)  E ,  
 \frac 1{X(\alpha)} 
\big[  j+ \frac 12 \nabla \alpha - X(\alpha) E\big] \Big\rangle
$$
Condition \eqref{c=} is equivalent to $X''(\alpha)=0$; in this case 
we can easily show, as before, that $\mc V$ is jointly convex.

\subsection{Periodic boundary conditions}
\label{sec5.7}

In this subsection we consider the case when $\Lambda = \bb T$, the
one dimensional torus of side length one and constant external field
$E$. If there is no external field, $E=0$, then it is an equilibrium
model; non equilibrium if $E\neq 0$.

In this case we have the possibility of constructing a space time path
$(\rho(t,u),j(t,u))$ of density and current in the form of a
travelling wave for the variational problem \eqref{limT} defining the
functional $\Phi$. We shall see that, under some assumptions on the
transport coefficient $D(\rho),\chi(\rho)$, this strategy for large
$J$ is more convenient than taking a density path $\rho$ constant in
time, so that $\Phi(J) <U(J)$.  \medskip

In the context of periodic boundary conditions, the proof that $\Phi =
U$ presented in the previous section applies if $D$ is constant.
In other words, we have that $\Phi (J) = U(J)$ for all $J$ provided
$D$ is constant, $\chi$ is concave and $E=0$.
\medskip

Let $\mc M_m(\bb T)$ the convex set of positive functions $\rho$ 
on the torus $\bb T$ such that $\int_0^1\!du \, \rho(u)=m$; we call
$m$ the mass of $\rho$. In this context,
\begin{equation*}
 U(J) \;=\; \inf_\rho \frac 12 \int_0^1 du\, \frac{\{J 
- J (\rho)\}^2} {\chi(\rho(u))} \;,
\end{equation*}
where the infimum is carried over $\mc M_m(\bb T)$ and $J(\rho)$ was
defined in \eqref{f8}. For each $v\in \bb R$, let $\Psi_v:
\bb R \to \bb R_+$ be defined by
\begin{equation}
\label{PSI}
\Psi_v (J) \;=\; \inf_\rho \frac 12 \int_0^1 du\, \frac{\{J + 
v[\rho(u) - m] - J (\rho)\}^2} {\chi(\rho(u))} \;,
\end{equation}
where the infimum is carried over $\mc M_m(\bb T)$.

We claim that 
\begin{equation}
\label{g02}
\Phi \;\le\; \Psi_v
\end{equation}
for each $v$. Indeed, consider a profile $\rho_0$ in $\mc M_m(\bb T)$.
Let $T=v^{-1}$ and set $\rho(t,u) = \rho_0(u-vt)$, $j(t,u)= J + v
[\rho_0(u-tv) -m]$ in the time interval $[0,T]$. An elementary
computation shows that the continuity equation holds and that the time
average over the time interval $[0,T]$ of $j(\cdot, u)$ is equal $J$.
In particular,
\begin{equation*}
\Phi(J) \;\le\; \frac 1T \int_0^T dt \, \mc U(\rho (t), j (t))\; .
\end{equation*}
On the other hand, it is easy to show by periodicity that the right
hand side is equal to
\begin{equation*}
\frac 12 \int_0^1 du\, \frac{\{J + 
v[\rho_0(u) - m] - J (\rho_0)\}^2} {\chi(\rho_0(u))} \;.
\end{equation*}
Optimizing over the profile $\rho_0$, we conclude the proof of
\eqref{g02}. 
\medskip

Fix a mass $m$, an external field $E$ and a current $J$. 
If $J^2/\chi + E^2 \chi$ is a convex function then 
\begin{equation}
\label{g01}
U(J) \;=\; \frac 12 \frac{\{J - E \chi(m)\}^2}{\chi(m)}
\end{equation}
and the optimal profile for the variational problem defining
$U(J)$ is the constant profile $\rho(u) =m $. 
In particular if $1/\chi$ and $E^2\chi$ are convex functions then
\eqref{g01} holds for any $J$ so that $U$ is trivially convex.

Indeed, fix a mass $m$, a current $J$ and an external field $E$. For
any profile $\rho$ in $\mc M_m(\bb T)$,
\begin{equation*}
\int_0^1 du\, \frac{\{J - J(\rho)\}^2}{\chi(\rho)}
\;=\; \int_0^1 du \, \frac{\{J - E \chi(\rho)\} ^2}{\chi(\rho)} 
\;+\; \int_0^1 du \, \frac{[(1/2) \nabla d(\rho)]^2}{\chi(\rho)} 
\end{equation*}
because the cross term vanishes upon integration 
Here $d(\rho)$ is the function introduced in the previous Subsection.
We have
$$
\int_0^1 \! du \, \frac{\{J - E \chi(\rho)\} ^2}{\chi(\rho)} 
= \int_0^1 \!du \, \Big[  \frac{J^2} {\chi(\rho)} + E^2 \chi(\rho)
  \Big]  -2 E J
\ge \frac{J^2} {\chi(m)} + E^2 \chi(m ) -2 E J
$$
where we used Jensen inequality, the convexity of $J^2/\chi + E^2
\chi$, and $\rho\in  \mc M_m(\bb T)$.

Therefore, for
all profiles $\rho$ in $\mc M_m(\bb T)$,
\begin{equation*}
\int_0^1 du \, \frac{[J - J(\rho)]^2}{\chi(\rho)}
\;\ge \; \frac{\{J - E \chi(m)\}^2}{\chi(m)} \;\cdot
\end{equation*}
Since the cost of the constant profile $\rho(u) = m$ is $(1/2) \{J - E
\chi(m)\}^2/\chi(m)$, \eqref{g01} is proven.

For the KMP model \cite{bgl, kmp} we have $D(\rho)=1$ and $\chi(\rho)
= \rho^2$. Since $\chi$ and $1/\chi$ are convex functions,  
the assumptions for \eqref{g01} are satisfied for any external field $E$. 
For the simple exclusion process we have  $D(\rho) =1$ and $\chi(\rho)
= \rho(1-\rho)$. In the case of no external field, $E=0$, since $\chi$
is concave and $\chi^{-1}$ is convex, it satisfies
both the hypotheses for $\Phi=U$ and the ones for \eqref{g01}; hence 
$\Phi(J) = (1/2) J^2/ m(1-m)$.   
If $E\neq 0$ the assumptions for \eqref{g01} holds only if $|E/J|$ is
small enough.

\bigskip
Fix a mass $m$, $e\in \bb R$ and take the external field $E= e J$.
Assume that 
\begin{equation}
\label{g03}
(1 - e^2 \chi(m)^2 ) \chi''(m) \; >\; 0
\end{equation}
We claim that there exists $w$ in $\bb R$ such that
\begin{equation}
\label{PIPPO}
\limsup_{|J|\to\infty} \frac{\Psi_{wJ} (J)}{J^2} 
<  \frac{\{1 - e \chi(m)\}^2}{2 \chi(m)} \;\cdot
\end{equation}
where $\Psi_v$ has been defined in \eqref{PSI}. 

Fix a mass $m$, a current $J$, an external field $E=eJ$ and take
$v=wJ$. For  $\mc M_m(\bb T)$,  by expanding the square we get
that
\begin{eqnarray}
\label{g04}
\!\!\!\!\!\!\!\!\!\!\!\! &&
\int_0^1 du\, \frac{\{J + 
wJ [\rho - m] + (1/2)\nabla d (\rho) - E \chi(\rho)\}^2}
{\chi(\rho)} \\
\!\!\!\!\!\!\!\!\!\!\!\! && \quad
=\; J^2 \int_0^1 du\, \frac{\{ 1 + w [\rho -m] - e \chi(\rho) \}^2}
{\chi(\rho)} \; +\;
\frac 14 \int_0^1 du\, \frac{[\nabla d(\rho)]^2}
{\chi(\rho)}\; \cdot 
\nonumber
\end{eqnarray}
because the cross term vanishes.  Expand the square on the first
integral. Let $F(r) = F_{w,m}(r)$ be the smooth function defined by
\begin{equation*}
F(r) = \frac{\{ 1 + w [r -m] \}^2}{\chi(r)} 
\; -2 e\; \;+\; e^2 \chi(r) \;\cdot
\end{equation*}
An elementary computation shows that 
\begin{equation*}
F''(m) = \frac 1{\chi(m)^3} \Big\{ 2 \chi(m)^2 w^2
- 4 \chi(m) \chi'(m) w + 2 \chi'(m)^2 - \chi(m) \chi''(m) 
+ e^2 \chi''(m) \chi(m)^3\Big\}\;.
\end{equation*}
Let $w= \chi'(m)/\chi(m)$. For this choice $F''(m)<0$. In particular,
we can choose a non constant profile $\rho (u)$ in $\mc M_m(\bb T)$
close to $m$ such that $F''(\rho(u)) <0$ for every $u$. Hence, by
Jensen inequality, the coefficient of $J^2$ in \eqref{g04} is strictly
less than
\begin{equation*}
\frac{\{ 1 - e \chi(m) \}^2}{\chi(m)} \;\cdot
\end{equation*}
The statement follows.

\medskip
Fix $e$, assume that $1/\chi + e^2 \chi$ is convex and
\eqref{g03} holds. Take $E=e J$.
Then, by \eqref{g02}, \eqref{g01} and \eqref{PIPPO} we have 
$\Phi(J) < U (J)$ for all sufficiently large currents $J$. 
The KMP model satifies the above requirements if $|e|< 1/m^2$.

The exclusion process with external field $E$ satisfies \eqref{g03}
for $|E|> |J|/m(1-m)$ but, for these values of $E$ and $J$ the function 
$1/\chi + e^2 \chi$ is not convex so we do not know whether $U$ is
given by \eqref{g01}. By  \eqref{PIPPO} we have that, 
for large current $J$, there exists a travelling wave whose cost is
strictly less than the one of the constant profile $\rho(u) = m$. 
This is not enough to prove the strict inequality $\Phi(J) < U(J)$.

We conclude by giving, for the KMP process with no external field, a
description of the mechanism for the strict inequality $\Phi<U$ in
terms of the power necessary, according to \eqref{lldrn}, to substain
a time average current $J$.
To get $U(J)= (1/2) J^2/m^2$ we switch on a constant external field
equal to $J/m^2$ which provides exactly the power $U(J)$.  On the
other hand we can impose a time average current $J$ by imposing a
space time dependent external field of the type $F(u-vt)$; the
corresponding density and current paths are then travelling waves. By
exploiting the convexity of $\rho^2$ we have shown that the second
strategy, at least for $J$ large, requires less power.

\subsection{An example with non convex $U$}
\label{sec5.2}

We discuss here a special choice of the macroscopic transport
coefficients $D$ and $\chi$ for which the functional $U$ defined in
\eqref{U} is not convex. In particular the upper bound \eqref{ub} with
the convex envelope differs from \eqref{1ub}.

We take $d=1$, $\Lambda= (0,1)$, $E=0$, $D(\rho)=1$, and $\chi(\rho)$
a smooth function with $\chi(0)=\chi(1)=0$ (accordingly the density
satisfies $0\le \rho \le 1$) such that there exist $0 < A < B <1$,
$\ell\in \bb R$ for which $\chi(\rho)= e^{-\ell \rho}$ if $A\le \rho
\le B$. Furthermore we take the equilibrium boundary conditions
$\rho(0)=\rho(1) = \bar\rho= (A+B)/2$.  We show that, for a suitable
choice of the parameters $A,B,\ell$, there are $J_1< J_2$ so that
$U''(J) < 0$ for any $J\in (J_1,J_2)$.

Although we did not construct explicitly a microscopic lattice gas
model in which the macroscopic transport coefficients meet the above
requirements, we believe it would be possible to 
exhibit a model which has the same qualitative behavior.  

For $d=1$, $D=1$, $E=0$, and $J\in \bb R$, the Euler--Lagrange
equation for the variational problem \eqref{U} defining the functional
$U(J)$ is
\begin{equation}
\label{el1}
\left\{
\begin{array}{l}
{\displaystyle
\frac 12 \rho''(u) = 
- \frac{\chi'(\rho(u))}{\chi(\rho(u))}
\big[ J^2 - \frac 14 \rho'(u)^2 \big]
\,,\qquad u\in (0,1)
}
\\
{\displaystyle
\rho(0) = \rho_0 \,,\qquad \rho(1) =\rho_1 
}
\end{array}
\right.
\end{equation}

For the above choice of $\chi$ and of the boundary conditions,
provided $4|J|\le B-A$, a solution of \eqref{el1} is given by
\begin{equation}
\label{alberto}
\hat\rho_J (u) = \bar\rho + \frac{2}\ell 
\log \frac{\cosh [J\ell(u-1/2)]}{\cosh [J\ell/2]}
\end{equation}
as can be easily verified. Note indeed that $A\le \hat\rho_J \le B$ 
since we assumed $4|J|\le B-A$.

We shall later prove that, under the above conditions,
\eqref{alberto} is the unique solution of the  boundary value problem
\eqref{el1}. A simple computation then gives 
\begin{equation*}
U(J) = \frac{2 e^{ \ell \bar\rho} }{ \ell^2} 
\, \frac{J \ell}{2} \tanh  \frac{J \ell}{2}
\end{equation*}
Let $F(z) := z \tanh z$ and $z^*$ be the unique positive root of 
$z^{-1}= \tanh z$. Then 
$F''(z) = 2 (1-\tanh^2 z)( 1 -z\tanh z) < 0$ 
for $z>z^*$. Hence $U''(J) <0$ if 
$ |J| \in \big( 2 z^*/\ell, (B-A)/4 \big)$. This interval is not empty
provided $\ell$ is chosen large enough.   

To show that \eqref{alberto} is the unique solution of the boundary
value problem \eqref{el1}, let us first prove that, given $J\neq 0$,
any solution of \eqref{el1} satisfies the \emph{a priori bound}
$|\rho'| \le 2|J|$. Since $\rho(0)=\rho(1)$, we can exclude the
possibility that $|\rho'(u)|\ge 2|J|$ for every $u\in[0,1]$.
By the continuity of $\rho'$, it is therefore enough to prove that 
$|\rho'(u)|\neq 2|J|$ for every $u\in[0,1]$. 
Suppose conversely that there exists $u^*\in[0,1]$ such that
$\rho'(u^*)= 2 J$. Then, by the uniqueness of the Cauchy problem
$\frac 12 \rho''=  \frac{\chi'(\rho)}{\chi(\rho)}\big[ J^2 - \frac 14 
\rho'^2 \big] $, $\rho(u^*)= \rho^*$, $\rho'(u^*)= 2 J$, we
would get that the solution of \eqref{el1} is $\rho(u) = \rho(u^*) + 2
J (u-u^*)$. Since this function does not satisfy the boundary
conditions in \eqref{el1} we find the desired contradiction.

Since $4|J| \le B-A$, the \emph{a priori bound} $|\rho'|\le 2|J|$ implies
that any solution $\rho$ of \eqref{el1} satisfies $A\le \rho\le
B$. For such values we have that $\chi'(\rho)/\chi(\rho) = -\ell$.
Uniqueness of the solution to \eqref{el1} can then be easily proven by
explicit computations.

\subsection{Zero range processes}
\label{sec4}

In this section we consider the so-called one-dimensional zero-range
processes which models a non-linear diffusion of lattice gases
\cite{kl} under constant external field $E$. The model is described by
positive integer-valued variables $\eta_x$ representing the number of
particles at site $x$. The particles jump with rates $(1/2) g(\eta_x))
\exp\{ E/N\}$ to right, $(1/2) g(\eta_x))$ $\exp\{- E/N\}$ to the left,
respectively. The function $g(k)$ is such that $g(k+1) - g(k) \ge a$
for some $a>0$ and $g(0)=0$. We assume that our system interacts with
particle reservoirs at the sites $0$ and $N$ whose activity is given
by $\varphi_0$, $\varphi_1$.

The generator of this Markov process is given by \eqref{genmc} with
$\Lambda_N = \{1, \dots, N\}$ and 
\begin{eqnarray*}
c_{x,x+1}(\eta) = g(\eta_x) \, e^{E/N} \; ,\quad 
c_{x+1,x}(\eta) = g(\eta_{x+1}) \, e^{-E/N}
\end{eqnarray*}
for $1\le x\le N-1$. Moreover, at the boundary,
\begin{eqnarray*}
\!\!\!\!\!\!\!\!\!\!\!\!\! &&
c_{0,1}(\eta) = \varphi_0 \, e^{E/N} \; ,\quad 
c_{1,0}(\eta) = g(\eta_1) \, e^{-E/N} \; ,\\
\!\!\!\!\!\!\!\!\!\!\!\!\! && \qquad 
c_{N,N+1}(\eta) = g(\eta_{N}) \, e^{E/N} \; ,\quad 
c_{N+1,N}(\eta) = \varphi_1 \, e^{-E/N} \;.
\end{eqnarray*}

Let $V(u) = E u$ and let $\varphi_N(x)$ be the solution of
$$
e^V \Delta_N \frac{\varphi_N}{e^V}  = \frac{\varphi_N}{e^V}
\Delta_N e^V
$$ 
for $1\le x\le N$ and with boundary condition $\varphi_N(0)
=\varphi_0$, $\varphi_{N}(N+1)= \varphi_1$. Here $\Delta_N$ stands for
the discrete Laplacian. 

The invariant measure $\mu_N$ is the grand--canonical measure
$\mu_N=\prod_{x\in\Lambda_N} \mu_{x,N}$ obtained by taking the product
of the marginal distributions
\begin{equation*}
\mu_{x,N}
(\eta_x = k) = \frac {1}{Z(\varphi_N(x))} \;
{\frac {\varphi_N (x)^k}{g(1)\cdots g(k)}}
\label{INV}
\end{equation*}
where $Z(\varphi) = \sum_{k\ge 0} \varphi^k/ [g(1) \cdots g(k)]$ is
the normalization constant. Let $R(\varphi) = \varphi Z'(\varphi)/$
$Z(\varphi)$ and denote by $\Psi$ its inverse function. 
For this process, the hydrodynamic equation (\ref{H1}) and the large
deviations principle (\ref{LD}) can be obtained with $D(\rho) =
\Psi'(\rho)$, $\chi (\rho) = \Psi(\rho)$, see \cite{BDGJL2}.
Moreover, $F_0'(\rho) =\log\Psi(\rho)$.

Let first show how, for this model, it is possible to solve
explicitly the variational problem (\ref{Uh}) for the Laplace
transform of the total current.  We note that the case $E=0$ has
already been solved in \cite{bd}.  Since we are in one space
dimension, the condition $\nabla \cdot \lambda = 0$ simply states that
$\lambda$ is a constant.  Moreover we have $J(\rho)= - (1/2)
\Psi'(\rho) \rho' + \Psi(\rho) E$, where hereafter the apices denotes
differentiation w.r.t.\ the macroscopic variable $u$.  Note that
condition \eqref{c=} holds so that $\Phi = U$. Changing variables in
(\ref{Uh}) by introducing $\varphi(u)= \Psi(\rho(u))$, $u\in[0,1]$; we
get
\begin{equation*}
U^*(\lambda) =
\frac 12  \sup_{\varphi} 
\int_0^1\!du\:  \Big\{ - \varphi(u) \psi'(u)^2
+ \lambda^2 \varphi(u)  
- \lambda  \varphi'(u) + 2 \lambda  E  \varphi(u) 
\Big\}
\end{equation*}
where the supremum is over all positive $\varphi$ such that 
$\varphi(0)=\varphi_0$, $\varphi(1)=\varphi_1$ and $\psi$ solves 
$$
2 \Big( \varphi(u) \psi'(u) \Big)' = \Big(  - \varphi'(u) +
2 [E+\lambda]  \varphi(u) \Big)'
$$
with boundary conditions $\psi(0)=\psi(1)=0$. The solution is 
$$
\psi(u) = - \frac 12 \log \frac{\varphi(u)}{\varphi_0} + (E+\lambda) u 
+ A \int_0^u\!dv \: \frac{1}{\varphi(v)}
$$
where
$$
A= \Big\{ \frac 12 \log \frac{\varphi_1}{\varphi_0} - (E+\lambda)\Big\}
\Big\{ \int_0^1\!du \: \frac{1}{\varphi(u)} \Big\}^{-1}
$$
After elementary manipulations, the variational problem for $U^*
(\lambda)$ becomes
\begin{equation*}
\begin{array}{l}
{\displaystyle 
U^*(\lambda) 
= \frac 12 \sup_{\varphi} \bigg\{
\Big\{- \frac 12 \log \frac{\varphi_1}{\varphi_0} + E+\lambda
  \Big\}^2 
\Big\{\int_0^1\!du \: \frac{1}{\varphi(u)} \Big\}^{-1}
}
\\ \qquad\qquad\qquad\qquad\qquad
{\displaystyle 
- \frac 14 \int_0^1\!du\: 
\frac{\varphi'(u)^2}{\varphi(u)}
- E^2 \int_0^1\!du \: \varphi(u)
+ E (\varphi_1 -\varphi_0)
\bigg\}
}
\end{array}
\end{equation*}
The associated extremality condition, which determines the optimal
profile, is  
\begin{equation}
\label{ec}
2 \varphi''(u) \varphi(u) - \big( \varphi'(u) \big)^2 - 4 E^2 
\varphi(u)^2  
= - 4 
\Big(- \frac 12 \log \frac{\varphi_1}{\varphi_0} + E+\lambda
  \Big)^2 \; 
\bigg[ \int_0^1\!dv \: \frac{1}{\varphi(v)}\bigg]^{-2} 
\end{equation}
with the boundary condition $\varphi(0)=\varphi_0$, $\varphi(1)=\varphi_1$.

In the case $E=0$ it is not difficult to check that the solution of
(\ref{ec}) is
$$
\varphi(u) =  C \Big( u + \frac{ e^{-\lambda}}{1- e^{-\lambda}} \Big)
\Big(u - \frac{ \varphi_0 e^{\lambda}}{\varphi_0 e^{\lambda} -
  \varphi_1} \Big)
$$
where $C= - \big( 1- e^{-\lambda}\big) \big( \varphi_0 e^\lambda
-\varphi_1\big)$. We then get
\begin{equation*}
U^* (\lambda) =
- \frac 14 [ \varphi'(1) -\varphi'(0)]
= \frac 12 
\big( 1- e^{-\lambda}\big) \big( \varphi_0 e^\lambda-\varphi_1\big)
\end{equation*}

In the case $E\neq 0$, the solution of (\ref{ec}) is instead given by 
$$
\varphi(u) =  C \Big( e^{2E u} -a \Big)\Big( e^{-2 E u} - b \Big)
$$
where 
$$
a = \frac{\varphi_0 e^{2E +\lambda} - \varphi_1}
{\varphi_0 e^{\lambda} - \varphi_1}\;\, \quad
b = \frac{1-e^{-\lambda - 2E}}{1-e^{-\lambda}}
$$
and
$$
C= \frac{(1-e^{-\lambda})(\varphi_0 e^{\lambda} - \varphi_1)}
{(e^{2E} - 1)(1-e^{-2E})}\;\cdot
$$ Notice that this solution converges, as $E\to 0$, to the solution
with no external field.  Plugging this solution into the variational
formula for $U^*$, we get that
\begin{equation}
\label{fa2}
U^*(\lambda) \;=\; - \frac 14 \, [ \varphi'(1) -\varphi'(0)]
\;+\; \frac 12 \, [ \varphi_1 -\varphi_0]
\;=\; E  \Big\{ \frac{\varphi_0}{1-e^{-2E}} (e^\lambda -1) 
+ \frac{\varphi_1}{e^{2E} -1}(e^{-\lambda}-1) \Big\} 
\; .
\end{equation}

We conclude this section showing that we may invert the order of
limits in \eqref{fa1} for zero range models.  For $0\le x, y\le N+1$,
$|x-y|=1$, recall that we denote by $\mc N^{x,y}_t$ the total number
of jumps from $x$ to $y$ in the time interval $[0,t]$. For $0\le x\le
N$, let $Q^{x,x+1}_t = \mc N^{x,x+1}_t - \mc N^{x+1,x}_t$ be the total
current over the bond $(x,x+1)$. Note that we are including the
boundary bonds.  

Consider the limit as microscopic time $t$ goes
to infinity of the Laplace transform of the total current:
$$
e_N(\lambda)\;=\; \frac 1N 
\lim_{t\to\infty} \frac 1t \log \bb E^N_{\eta^N}
\Big [\exp\Big\{  \lambda N^{-1} \sum_{x=0}^{N} Q^{x,x+1}_t \Big\} \Big]
$$ 
and notice that $f_N$ given by (\ref{f5}) is related to the
Legendre transform of $e_N$ by
\begin{equation*}
f_N^* (\lambda) \;=\; e_N \Big( \lambda \big\{ 2E - \log
(\varphi_1/\varphi_0)\big\} \Big)\;.
\end{equation*}
Notice furthermore that this expression does not depend on the initial
condition $\eta^N$ by ergodicity.

Since two currents $Q^{x,x+1}_t$ $Q^{y,y+1}_t$ differ only by surface
terms, in the asymptotic $t\uparrow\infty$, we may replace all
currents by $Q^{0,1}_t$ and obtain that
\begin{equation*}
\label{f7}
e_N(\lambda) \;=\; \frac 1N
\lim_{t\to\infty} \frac 1t \log \bb E_\mu \big [e^{ \lambda
Q^{0,1}_t }\big]\;.
\end{equation*}

To compute the previous limit, we represent the zero range process in
terms of interacting random walks.  Let $N_0$ be the total number of
particles at time $0$: $N_0 = \sum_x \eta_x(0)$. We start labeling
these particles. New particles entering the system at the boundary get
new labels in an increasing order. Denote by $X^i(t)$ the position at
time $t$ of the $i$-th particle. $X^1$ performs a weakly asymmetric
random walk on $\Lambda_N$ with absorption at the boundary and mean
$g(1)$ exponential waiting times. $X^2$ does the same but its clock
rates are affected by $X^1$. If they occupy different sites, the
$X^2$-exponential has rate $g(1)$, while if both occupy the same site,
its exponential clock has rate $g(2)-g(1)$ and so on. We need for this
construction the function $g$ to be increasing. Moreover the condition
$g(k+1)-g(k)\ge a > 0$ guarantees that these random walks will hit the
boundary with probability one.

Let $w_i$ (resp.\ $u_i$) the indicator function of the event that the
$i$-th particle created at the left (resp.\ right) boundary is absorbed
at $N+1$ (resp.\ $0$).  Denote by $N_\pm(t)$ a Poisson process of rate
$\varphi_0/2$, $\varphi_1/2$ which represents the entrance of
particles at either boundary. With this notation, up to negligible
terms in the limit $t\uparrow\infty$,
$$
Q^{0,1}_t \;=\; \sum_{i=1}^{N_{-}(t)} w_i  - \sum_{i=1}^{N_+(t)} u_i \;.
$$
Since the random variables $w_i$, $u_j$ are independent, elementary
computation shows that
$$
e_N(\lambda) = \frac 1N \Big\{ (\varphi_0/2) p_N \{ e^\lambda -1 \}
+ (\varphi_1/2) q_N \{ e^{-\lambda} -1 \} \Big\} \;,
$$ 
where $p_N = P[w_i=1]$ (resp.\ $1-q_N$) is the probability that a
random walk, absorbed in $0$ and $N+1$, with transition probability
$p(x,x+1) = e^{2 E/N} / (e^{2E/N} +1) = 1-p(x,x-1)$ starting from $1$
(resp.\ $N$) is absorbed in $N+1$ (resp.\ $0$). These probabilities can
be explicitly computed.

As $N\uparrow\infty$, we get that
\begin{eqnarray*}
\lim_{N\to\infty} e_N(\lambda)\;=\; E
\Big\{ \frac{\varphi_0}{1-e^{-2E}}
\big(e^{\lambda } -1 \big)
+  \frac{\varphi_1}{e^{2E}-1}
\big(  e^{-\lambda} -1 \big) \Big\}
\end{eqnarray*}
which agrees with \eqref{fa2}.

\appendix

\Section{Supplement to Section~\protect\ref{sec1}.}
\label{secap}
\def\theequation{A.\arabic{equation}}

We present here a derivation, at the heuristic level, of the
law of large numbers and the large deviations, as $N\to\infty$, for
the empirical density and the empirical current. 
Recall the notation introduced in Section \ref{sec1}.  We have seen
there that to prove the law of large numbers for the empirical measure
and the current, we need to express the limit current $J$ in terms of
the density $\rho$.

In the context of stochastic lattice gases this is done by assuming a
\emph{local equilibrium state}. Roughly speaking, this means that in a
large microscopic region $\Delta$ around $u$, still infinitesimal
macroscopically, the system has relaxed to the Gibbs state (with
Hamiltonian $\mc H$) conditioned to $\sum_{x\in\Delta}\eta_x= |\Delta|
\pi^N(t,u)$.  This assumption, which can be rigorously justified, 
\cite{kl}, allows us to express the empirical current in terms of the
empirical density. We next show how this can be done for the
so--called \emph{gradient} models.

By standard computations in the theory of Markov processes we have
that, \cite[Lemma II.2.3]{sp}, for a bond $\{x,x+e_j\}$, 
\begin{equation*}
Q^{x,x+e_j}(t) \;=\; (1/2) N^2 \int_0^t\! ds \,
\big[ c_{x,x+e_j}(\eta_s) -c_{x+e_j,x}(\eta_s) \big]
\;+\; M^{x,x+e_j}(t)\;,
\end{equation*}
where $M^{x,x+e_j}(t)$ are martingales with bracket
\begin{equation*}
\langle M^{x,x+e_i}, M^{y,y+e_j} \rangle(t) = (1/2) N^2
\delta_{x,y} \delta_{i,j} 
\int_0^t\!ds \: 
\big[ c_{x,x+e_i}(\eta(s) ) +c_{x+e_i,x}(\eta(s)) \big]\;.
\end{equation*}
Let $G$ be a smooth vector field as in (\ref{empcurr}) vanishing on
$\partial\Lambda$. By definition of the martingales $M^{x,x+e_j}(t)$,
\begin{equation}
\label{3.7}
\langle\!\langle \mc J^N , G \rangle\!\rangle_T = 
\frac 12  \frac{1}{N^d}\int_0^T\!dt \sum_{i=1}^d \sum_x G_i(t,x/N) 
 N \big[ c_{x,x+e_i}(\eta(t)) -c_{x+e_i,x}(\eta(t)) \big]  
+ \mc M^N_T (G)\;, 
\end{equation}
where $\mc M^N_T (G)$ is a martingale term. An easy computation, based
on the explicit formula for the quadratic variations of the
martingales $M^{x,x+e_j}(t)$, shows that $\mc M^N_T (G)$ vanishes as
$N\to\infty$.  We next use definition (\ref{ratas}) and Taylor
expansion to write
\begin{equation}
\label{3.8}
\begin{array}{l}
{\displaystyle 
c_{x,x+e_i}(\eta) -c_{x+e_i,x}(\eta)
} \\
{\displaystyle 
\quad =
c^0_{x,x+e_i}(\eta)
\big[ 1 + \frac{1}{N} E_i (x/N) \big]
-c^0_{x+e_i,x}(\eta)
\big[ 1 -\frac{1}{N} E_i (x/N) \big] + O(1/N^2)
}
\\
\quad
{\displaystyle 
=\big[ c^0_{x,x+e_i}(\eta) - c^0_{x+e_i,x}(\eta) \big]
+ \frac{1}{N} 
\big[ c^0_{x,x+e_i}(\eta) + c^0_{x+e_i,x}(\eta) \big]
E_i  (x/N) + O(1/N^2)
}
\end{array}
\end{equation}
The \emph{gradient condition}, see \cite[II.2.4]{sp}, holds if there
exist local functions $h_0^{(i)}(\eta)$, $i=1,\dots,d$,  
depending on the configuration $\eta$ around $0$, so that for any
$i=1,\cdots, d$
\begin{equation}
\label{gc}
c^0_{x,x+e_i}(\eta) - c^0_{x+e_i,x}(\eta)
= h^{(i)}_{x} (\eta) - h^{(i)}_{x+e_i} (\eta) 
\end{equation}
where $h^{(i)}_x$ is the function $h_0^{(i)}$ evaluated on the
configuration $\eta$ translated by $x$.  Let us plug the right hand
side of (\ref{3.8}) into (\ref{3.7}).  By the gradient condition
(\ref{gc}) we can perform a summation by parts on the first term. Note
that there are no boundary terms since we assumed $G$ to vanish on the
boundary. We get, with a negligible error as $N\to\infty$,
\begin{equation}
\label{3.9}
\begin{array}{l}
{\displaystyle
\vphantom{\Bigg(}
\langle\!\langle \mc J^N , G \rangle\!\rangle_T \approx
\frac 12  \frac{1}{N^d}\int_0^T\!dt \sum_{i=1}^d \sum_x 
\Big\{ \partial_i G_i(t,x/N) h^{(i)}_x (\eta)
} \\ \qquad\qquad\qquad\quad
{\displaystyle
\vphantom{\Bigg(}
\phantom{\langle\!\langle \mc J^N , G \rangle\!\rangle_T}
+ G_i(t,x/N) 
\big[ c^0_{x,x+e_i}(\eta(t)) + c^0_{x+e_i,x}(\eta(t)) \big]  
E_i(x/N)
\Big\}
}
\end{array}
\end{equation}

Recall the definition of the functions $d^{(i)}$, $\chi^{(i)}$
introduced in \eqref{dc}.  By the local equilibrium assumption
mentioned above and the equivalence of ensembles from (\ref{3.9})
we get
\begin{equation}
\label{3.11}
\langle\!\langle \mc J^N , G \rangle\!\rangle_T \approx
\sum_{i=1}^d 
\int_0^T\!\!\!\!dt\! \int_\Lambda\!\!\!du 
\Big\{ \frac 12  \partial_i G_i(t,u) d^{(i)}\big(\pi^N(t,u)\big)
+ G_i(t,u)  \chi^{(i)} \big(\pi^N(t,u)\big) E_i(u)
\Big\}
\end{equation}
Taking the limit $N\to\infty$, the empirical density $\pi^N(t,u)$
converges to $\rho(t,u)$, whereas the empirical current $\mc J^N(t,u)$
converges to a vector field $J(t,u)$. Equation (\ref{3.11}) then 
implies
\begin{equation}
\label{f81}
J(\rho) = -\frac 12  D(\rho) \nabla \rho  + \chi(\rho) E 
\end{equation}
where $D$ and $\chi$ are $d\times d$ diagonal matrices with entries
$D_{ii}(\rho) =  \frac{d}{d\rho} d^{(i)} (\rho)$ and 
$\chi_{ii}(\rho) =  \chi^{(i)} (\rho)$. 
\medskip

We now turn to a heuristic derivation of the large deviations
principle (\ref{f1})--(\ref{Ic}) for the current. Recall the statement
and the notation introduced in Section \ref{sec1}.

In order to make the trajectory $j$ typical, we introduce an extra
weak time dependent external field $F=(F_1, \dots, F_d)$ by perturbing
the rates as in Section \ref{sec0}, namely
\begin{equation}
\label{rF}
c_{x,x+e_i}^F(\eta) =  c_{x,x+e_i}(\eta) \, e^{N^{-1} F_i(t, x/N)}\;,
\quad 
c_{x+e_i,x}^F(\eta) =  c_{x+e_i,x}(\eta) \, e^{- N^{-1} F_i(t, x/N)}\;.
\end{equation}
We denote by $\bb P^{N,F}_{\eta^N}$ the probability distribution of
the perturbed process. Since these rates $c^F$ are the same as the
rates of the original process with $E$ replaced by $E+F$
(cf. \eqref{ratas}), we have the following law of large numbers:
\begin{equation*}
\lim_{N\to\infty} 
\bb P^{N,F}_{\eta^N} \big( \mc J^N \approx j \big) = 1
\end{equation*}
where
\begin{equation}
\label{jEF}
j = J(\rho) + \chi(\rho) F = 
- \frac 12 D(\rho)\nabla \rho + \chi(\rho) ( E + F ) 
\end{equation}
and $\rho$ satisfies the continuity equation (\ref{ce}).

We now read this equation in the opposite direction: given the
trajectory $j$ we first solve (\ref{ce}) to get $\rho$, then we 
determine the external field $F$ which makes $j$ the
typical behavior, namely
\begin{equation}
\label{F=}
F =  \chi(\rho)^{-1} \big( j + \frac 12 D(\rho)\nabla \rho  \big) -E
\end{equation}

By writing the original process in terms of the perturbed one we have  
\begin{equation*}
\bb P_{\eta^N}^N 
\Big( \mc J^N (t,u) \approx j (t,u), 
\: (t,u) \in [0,T]\times \Lambda \Big) 
= \bb P^{N,F}_{\eta^N} \Big( 
\frac {d\bb P_{\eta^N}^N}{d\bb P_{\eta^N}^{N,F}}\, 
\id_{ \{\mc J^N \approx j\}}  \Big) 
\end{equation*}
The large deviation principle (\ref{f1})--(\ref{Ic}) will follow,
recalling (\ref{F=}), once we compute the Radon--Nikodym derivative and show 
that on the event $\{\mc J^N \approx j\}$ we have, 
with a negligible error if $N\to\infty$,
\begin{equation}
\label{lldrn}
\log \frac{d{\bb P}_{\eta^N}^{N}}{ d {\bb P}_{\eta^N}^{N,F} }
=
-\log \frac{d{\bb P}_{\eta^N}^{N,F}}{ d {\bb P}_{\eta^N}^{N} }
\approx  
- N^d  \frac 12  \int_0^T \!dt \: \langle F , \chi(\rho) F \rangle
\end{equation}
This equation can be interpreted, in analogy to the classical Ohm's
law, as the total work done in the time interval $[0,T]$ by the
external field $F$.

\medskip
We shall need some basic tool from the general theory of jump Markov
processes that we briefly recall, see e.g.\ \cite[Appendix A1]{kl} 
or  \cite[Appendix A]{BDGJL2}. 
Let $\Omega$ be a countable set and consider a continuous time jump 
Markov process $X_t$ on the state space $\Omega$ with generator given by
\begin{equation}
\label{genas}
L f (\eta) = \sum_{\eta'\in\Omega} \lambda(\eta) 
p(\eta,\eta') \left[ f(\eta') - f (\eta) \right]
\end{equation}
where the rate $\lambda$ is a positive function on $\Omega$ and
$p(\eta,\eta')$ is a transition probability. 
We consider also another process $X^F_t$ of the same type 
with time dependent rate $\lambda^F(\eta,t)$ and
transition probability $p^F(\eta,\eta';t)$. Then, denoting by 
$\bb P_{\eta_0} $ and $\bb P^F_{\eta_0}$ the distribution of the two processes
with initial condition $\eta_0$ we have
\begin{equation}\label{RNd}
\begin{array}{l}
{\displaystyle
\frac{d{ \bb P}^F_{\eta_0}}{d{ \bb P}_{\eta_0}} \big( 
X_t, \, t \in [0,T] \big)
}
\\ 
{\displaystyle \;\; = 
\exp\left\{ \sum_{i=1}^{n} \log 
\frac{\lambda^F(X_{\tau_{i-1}},\tau_i) 
p^F\left(X_{\tau_{i-1}}, X_{\tau_i}; \tau_i \right) }
{\lambda(X_{\tau_{i-1}}) 
p\left(X_{\tau_{i-1}}, X_{\tau_i}\right) }
- \int_0^T\! dt \,
\big[ \lambda^F(X_t,t) - \lambda(X_{t}) \big]
\right\}
}
\end{array}
\end{equation}
where $\tau_0=0$, $X_0=\eta_0$, $\tau_i$, $i=1,\dots,n$ is the time in 
which the process jumped from $X_{\tau_{i-1}}$ to $X_{\tau_i}$,
and $n$ is total number of jumps in the time interval $[0,T]$.  

For the process $\eta(t)$ with generator (\ref{genmc}) we have for
$x$, $y$ in $\bb Z^d$, 
\begin{equation*}
\lambda(\eta) \,  p(\eta,\sigma^{x,y} \eta ) \;=\;
(1/2) N^2 c_{x,y}(\eta) \; \qquad
\lambda(\eta) \;=\;
(1/2) N^2 \sum_{x,y} c_{x,y}(\eta) 
\end{equation*}
For the process $\eta(t)$ with rates (\ref{rF}) we have
\begin{equation*}
\lambda^F(\eta,t) \, p^F(\eta,\sigma^{x,x+e_i} \eta,t) \;=\;
(1/2) N^2 c_{x,x+e_i}(\eta) e^{N^{-1} F_i(t, x/N)}
\end{equation*}
and a similar formula for $\lambda^F(\eta,t) \, p^F(\eta,\sigma^{x+e_i,x}
\eta,t)$ so that
\begin{equation*}
\lambda^F(\eta) \;=\; (1/2) N^2 \sum_{i=1}^d \sum_{x} 
\Big\{ c_{x,x+e_i}(\eta) e^{N^{-1} F_i(t,x/N)} +
c_{x+e_i,x}(\eta) e^{- N^{-1} F_i(t,x/N)} \Big\}\;.
\end{equation*}
From (\ref{RNd}) and the explicit expressions for the rates, we get
that
\begin{eqnarray*}
\!\!\!\!\!\!\!\!\!\!\!\!\!\! &&
\log \frac{d{\bb P}_{\eta^N}^{N,F}}{ d{\bb P}_{\eta^N}^{N}}
= \frac 1N \sum_{j=1}^d \sum_{x} \Big\{ \sum_{\tau_{x,x+e_j}}  
F_j (\tau_{x,x+e_j},x/N) - \sum_{\tau_{x+e_j,x}} 
F_j (\tau_{x+e_j,x},x/N) \Big\} \\ 
\!\!\!\!\!\!\!\!\!\!\!\!\!\! && \qquad
- \frac{N^2}2  \int_0^T\! dt \, \Big\{ 
c_{x,x+e_j}( \eta (t)) \Big ( e^{N^{-1} F(t,x/N)} -1 \Big)
+ c_{x+e_j,x}( \eta (t)) \Big ( e^{- N^{-1} F(t,x/N)} -1 \Big)
\Big\}\;,
\end{eqnarray*}
where $\tau_{x,y}$ are the jump times from $x$ to $y$. Expanding the
exponentials and recalling the definition of the empirical current,
we may rewrite the previous expression as
\begin{equation}
\label{lrn}
\begin{array}{l}
{\displaystyle
N^d \langle\!\langle \mc J^N, F \rangle\!\rangle_T
\;-\; \frac N2 \int_0^T\!dt\: \sum_{x} \sum_{i=1}^d  
F_i(x/N,t) \, \big\{ c_{x,x+e_i}(\eta(t)) -  c_{x+e_i,x}(\eta(t)) \big\}
}
\\ \qquad
{\displaystyle
- \; \frac 12 \int_0^T\!dt\: \sum_{x} \sum_{i=1}^d \big\{ c_{x,x+e_i}(\eta(t)) 
+  c_{x+e_i,x}(\eta(t)) \big\} \, F_i(x/N,t)^2 \;+\; O(1/N) \;,
}
\end{array}
\end{equation}
where we let $c_{x,y} =0 $ if $x,y\not\in\Lambda_N$. 

For gradient models condition (\ref{gc}) holds so that we can perform
a summation by parts in the second term. Recalling the definition of
the diffusion matrix $D$, the mobility $\chi$ and the local
equilibrium assumption, we can express the second term of the right
hand side of (\ref{lrn}) in terms of the empirical density. Since we
are assuming $\mc J^N\approx j$, we get that
\begin{equation}
\log \frac{d{\bb P}_{\eta^N}^{N,F}}{ d{\bb P}_{\eta^N}^{N} }
\approx 
N^d \Big\{ \langle\!\langle j, F \rangle\!\rangle_T 
\;+\; \int_0^T\!dt  \langle F,  (1/2) D(\rho) \nabla \rho 
- \chi(\rho) E  \rangle 
-  \frac 12 \int_0^T\!dt \langle F, \chi(\rho) F \rangle 
\Big\}\;.
\end{equation}
which, by the choice of $F$ in (\ref{F=}), concludes the derivation of
(\ref{lldrn}).

The rigorous derivation of action functional $\mc I$ requires some
difficult estimates.  In fact, while in the proof of the hydrodynamic
limit it is enough to show that the local equilibrium assumption holds
with a negligible error as $N\to\infty$, in the proof of the large
deviations we need such an error to be $o(e^{-C N^d})$. This can be
proven by the so called super exponential estimate, see \cite{kl,kov},
which is the key point in the rigorous approach.
\medskip

Recall the dynamical large deviations principle for the density stated
in \eqref{LD}.  The rate functional $\mc F$ is given by
\begin{equation}
\label{rfden}
\mc F_{[0,T]}(\rho )\;=\; \frac 12 \int_0^T \!dt \,
\big\langle  \nabla H (t), \chi(\rho(t))  \nabla H(t)\big\rangle
\end{equation}
where, given the fluctuation $\rho$, 
the external potential $H=H(t,u)$ is chosen so that it vanishes at the 
boundary and 
\begin{equation}
\label{extper}
\partial_t \rho = 
\nabla \cdot \Big( \frac 12 D(\rho) \nabla\rho
- \chi(\rho) \big[ E + \nabla H \big] \Big)
\end{equation}
which is a Poisson equation for $H$.  

We conclude this Appendix showing how the above result follows
directly from the large deviation principle for the current.  We fix a
path $\rho=\rho(t,u)$, $ (t,u)\in[0,T]\times\Lambda$.  There are many
possible trajectories $j=j(t,u)$, differing by divergence free vector
fields, such that the continuity equation (\ref{ce}) is satisfied.
The functional $\mc F_{[0,T]}(\rho)$ can be obtained by minimizing
$\mc I_{[0,T]} (j)$ among all such paths $j$
\begin{equation}
\label{3.34}
\mc F_{[0,T]}(\rho) = 
\inf_{ \substack{ j \, : \\ \nabla \cdot j = -\partial_t \rho}} 
\mc I_{[0,T]} (j)
\end{equation}

To derive the functional (\ref{rfden}) we show that the
infimum above is obtained when the external perturbation $F$ introduced
in (\ref{F=}) is a gradient vector field whose potential $H$ solves 
(\ref{extper}).  Let $H$ be the solution of (\ref{extper}) and $F$ as
in (\ref{F=}), we write
\begin{equation}
F = \nabla H + \widetilde F
\end{equation}
By the definition of $H$  we get
$$
\langle \nabla H, \chi(\rho)  \widetilde F \rangle 
= - \Big\langle H,  
\nabla \cdot j  + \nabla \cdot \Big( \frac 12 D(\rho) \nabla\rho 
- \chi(\rho) E - \chi(\rho) \nabla H \Big) \Big\rangle
= 0
$$  
Hence
$$
\mc I_{[0,T]} (j) = \frac 12 \int_0^T\!dt \: 
\Big\{ \langle \nabla H, \chi(\rho)  \nabla H\rangle 
+ \langle \widetilde F, \chi(\rho)  \widetilde F \rangle 
\Big\}
$$
Therefore the infimum in (\ref{3.34}) is obtained when 
$\widetilde F =0$, so that the functional defined in  (\ref{3.34})
coincides with (\ref{rfden}).

\bigskip\bigskip

\noindent{\bf Acknowledgments.}
It is a pleasure to thank T.\ Bodineau, B.\ Derrida, G.\ Gallavotti,
E.\ Presutti, C.\ Toninelli, and S.R.S.\ Varadhan for stimulating discussions. 
The authors acknowledge the support of 
PRIN MIUR 2004028108   and 2004015228.

\end{document}